\documentclass[final]{l4dc2026}

\usepackage{graphics} 
\usepackage{times} 
\usepackage{amsfonts} 
\usepackage{xcolor}
\usepackage{cases}
\usepackage{cite}
\usepackage{algorithm}
\usepackage{algorithmic}
\usepackage[font=footnotesize]{caption}
\usepackage{subcaption}
\usepackage{pgfplots}
\usepackage{pgfplotstable}
\usepackage{tikz}
\usepackage{mathtools}
\usepgfplotslibrary{fillbetween}
\pgfplotsset{compat=1.16}
\usetikzlibrary{intersections}
\usetikzlibrary{patterns}
\usepackage{hyperref}

\usepackage{xurl}



\newtheorem{definition}{Definition}
\newtheorem{theorem}{Theorem}
\newtheorem{lemma}{Lemma}
\newtheorem{proposition}{Proposition}
\newtheorem{corollary}{Corollary}
\newtheorem{assumption}{Assumption}
\newtheorem{remark}{Remark}

\makeatletter
\newcommand{\blfootnote}[1]{%
  \begingroup
  \renewcommand{\thefootnote}{}%
  \renewcommand{\@makefntext}[1]{##1}
  \footnote{#1}%
  \addtocounter{footnote}{-1}%
  \endgroup
}
\makeatother

\usepackage{acronym}
\acrodef{pac}[PAC]{Probably Approximately Correct}

\title[Finite-Step Invariant Sets for Hybrid Systems with Probabilistic Guarantees]{Finite-Step Invariant Sets for Hybrid Systems with \\ Probabilistic Guarantees}
\usepackage{times}

\author{%
 \Name{Elizabeth Dietrich}$^{*}$ \Email{eadietri@berkeley.edu}\\
 \addr University of California, Berkeley%
 \AND
 \Name{Varun Madabushi}$^{*}$ \Email{vmadabushi@gatech.edu}\\
 \addr Georgia Institute of Technology%
 \AND
 \Name{Murat Arcak} \Email{arcak@berkeley.edu}\\
 \addr University of California, Berkeley%
 \AND
 \Name{Hanna Krasowski} \Email{krasowski@berkeley.edu}\\
 \addr University of California, Berkeley%
 \AND
 \Name{Maegan Tucker} \Email{mtucker@gatech.edu}\\
 \addr Georgia Institute of Technology%
 \vspace{-1.5cm}
}

\begin{document}
\maketitle
\blfootnote{$^{*}$Equal contribution, ordered alphabetically.}

\begin{abstract}%
Poincar\'e return maps are a fundamental tool for analyzing periodic orbits in hybrid dynamical systems by reducing the stability analysis of a periodic orbit to that of a discrete-time system.
While linearization provides local stability information, assessing robustness to disturbances requires identifying invariant sets of the state space under the return dynamics. 
However, computing such invariant sets is computationally difficult, especially when system dynamics are only available through forward simulation.
In this work, we propose an algorithmic framework leveraging sampling-based optimization to compute a finite-step invariant level set around a nominal periodic orbit using sampled evaluations of the return map.
The resulting solution is accompanied by probabilistic guarantees on finite-step invariance.
We provide theoretic analysis establishing sufficient conditions for which the algorithm terminates and demonstrate the approach on two low-dimensional systems and a Compass-Gait Walking model.
\end{abstract}

\begin{keywords}%
  Hybrid Systems, Probabilistic Guarantees, Invariance%
\end{keywords}
\section{Introduction}
Hybrid dynamical systems naturally arise in many robotic and cyber-physical systems that exhibit both continuous-time evolution and discrete events. 
Examples include robotic manipulation with intermittent contact \citep{kolathaya2018direct}, legged locomotion \citep{reher2021dynamic}, and cyber-physical systems, such as traffic networks \citep{coogan2015compartmental}.
The combination of discrete transitions and nonlinear, continuous dynamics introduces significant challenges in the analysis of hybrid systems compared to purely continuous systems.
A common tool for analyzing periodic behaviors in hybrid systems is the Poincar\'e return map, which reduces hybrid dynamics to a discrete-time system, defined on a guard surface.
In this framework, periodic behaviors correspond to fixed points of the return map, whose stability can be analyzed using standard discrete-time techniques.
This approach has been widely applied to the study of periodic motions in hybrid systems, including robotic locomotion and other systems with impacts \citep{grizzle2008hybrid, agrawal2017first, veer2019input, goodman2019existence, gong2022zero}.
However, local stability alone does not characterize trajectory deviations from the nominal orbit under disturbances and model uncertainty, motivating the need for a comprehensive notion of \emph{robustness}.

A natural way to characterize such robustness is through forward-invariant sets of the return map, which ensure that trajectories initialized within the set remain contained for all time.
These sets capture the perturbations that a system can tolerate while maintaining stable behavior.
In practice, computing invariant sets for hybrid systems is challenging.
Existing approaches either rely on explicit models, using sum-of-squares programming \citep{manchester2010regionsattractionhybridlimit}, Hamilton--Jacobi reachability \citep{choi2022regions}, or set propagation \citep{madabushi2026}, or learn a surrogate of the return map from sampled data \citep{werner2026halohybridautoencodedlocomotion}.
For learned tracking controllers trained in simulation \citep{janwani2026navigait, li2021robustparameterized}, explicit models of the controller and dynamics are unavailable, motivating the use of data-driven approaches.
However, existing data-driven methods only provide guarantees with respect to a learned surrogate model of the dynamics \citep{werner2026halohybridautoencodedlocomotion} or restrict the invariant set to a uniform ball centered at the nominal gait \citep{tucker2024synthesizing}.

\begin{figure}
    \centering
\usepgfplotslibrary{groupplots}
\pgfplotsset{compat=1.18}
\usetikzlibrary{positioning,arrows.meta,calc}
\graphicspath{{l4dc_figures/pdfs/}}

\definecolor{cW}{HTML}{5302A3}
\definecolor{cU}{HTML}{B12A90}
\definecolor{cZ}{HTML}{E16462}
\definecolor{cV}{HTML}{FCA636}
\definecolor{cXi}{HTML}{AAAAAA}
\definecolor{cPrev}{HTML}{888888}

\newcommand{\dotleg}[1]{\tikz[baseline=-0.6ex]\fill[#1] (0,0) circle (2.2pt);}
\newcommand{\ellleg}{\tikz[baseline=-0.6ex]\draw[black,thick] (0,0) ellipse (4pt and 2.5pt);}
\newcommand{\prevleg}{\tikz[baseline=-0.6ex]\draw[cPrev,thick,dotted] (0,0) ellipse (4pt and 2.5pt);}

\newcommand{\axleg}[2]{%
  \node[anchor=south, font=\scriptsize, align=left, inner sep=2pt,
        fill=white, fill opacity=0.85, text opacity=1,
        draw=black!20, rounded corners=1pt]
    at ([xshift=0mm,yshift=0mm]#1.south) {#2};}
    
\def\xmin{-2.5}\def\xmax{2.5}
\def\ymin{-1.5}\def\ymax{2.5}

\noindent\hspace*{-.2cm}%
\resizebox{\linewidth}{!}{%
\begin{tikzpicture}
\begin{groupplot}[
    group style={
      group size=4 by 1,
      horizontal sep=0.5cm,
      yticklabels at=edge left,
      ylabels at=edge left,
    },
    width=0.25\linewidth, height=0.25\linewidth, scale only axis,
    xmin=\xmin, xmax=\xmax, ymin=\ymin, ymax=\ymax,
    xtick={-2,0,2}, ytick={-1,0,1, 2},
    tick label style={font=\large},
    label style={font=\large},
    title style={font=\normalsize, yshift=-2pt},
    xlabel={$x_1$}, ylabel={$x_2$},
    axis on top,
]
\nextgroupplot[title={(a) Step 1: Sample from $\hat{\mathcal{I}}(\theta_k)$}]
  \addplot graphics[xmin=\xmin,xmax=\xmax,ymin=\ymin,ymax=\ymax]{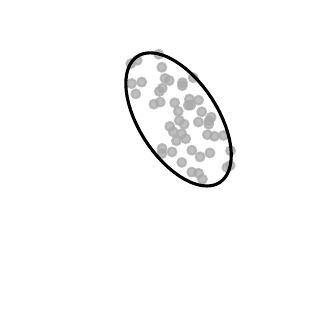};
\nextgroupplot[title={(b) Step 2: Apply $\mathcal{P}$}]
  \addplot graphics[xmin=\xmin,xmax=\xmax,ymin=\ymin,ymax=\ymax]{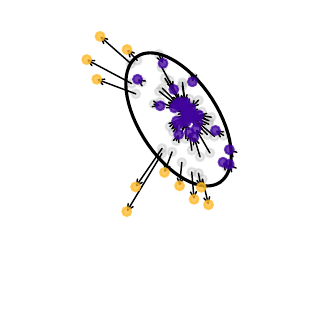};
\nextgroupplot[title={(c) Step 3: Partition}]
  \addplot graphics[xmin=\xmin,xmax=\xmax,ymin=\ymin,ymax=\ymax]{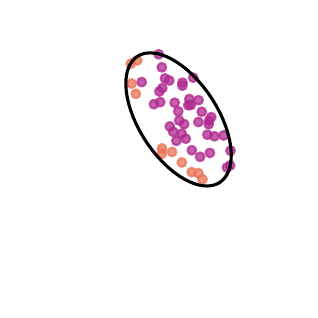};
\nextgroupplot[title={(d) Step 4: Fit to $\mathcal{U}$}]
  \addplot graphics[xmin=\xmin,xmax=\xmax,ymin=\ymin,ymax=\ymax]{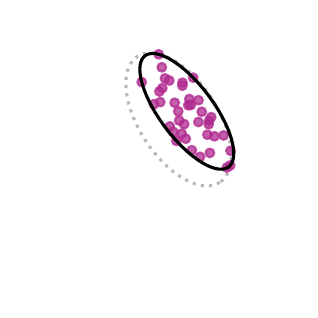};
\end{groupplot}

\axleg{group c1r1}{\ellleg\ $\hat{\mathcal{I}}(\theta_k)$ $\quad $
                   \dotleg{cXi}\ $\{x_i\}$}

\axleg{group c2r1}{\dotleg{cW}\ $\mathcal{W}_k:=\{\mathcal{P}(x_i)\mid \mathcal{P}(x_i)\in\hat{\mathcal{I}}(\theta_k)\}$\\
                   \dotleg{cV}\ $\mathcal{V}_k:=\{\mathcal{P}(x_i)\mid \mathcal{P}(x_i)\notin\hat{\mathcal{I}}(\theta_k)\}$}

\axleg{group c3r1}{\dotleg{cU}\ $\mathcal{U}_k:=\{x_i\mid \mathcal{P}(x_i)\in\mathcal{W}_k\}$\\
                   \dotleg{cZ}\ $\mathcal{Z}_k:=\{x_i\mid \mathcal{P}(x_i)\in\mathcal{V}_k\}$}

\axleg{group c4r1}{\prevleg\ $\hat{\mathcal{I}}(\theta_k)$ $\quad$
                   \ellleg\ $\hat{\mathcal{I}}(\theta_{k+1})$ $\quad$
                   \dotleg{cU}\ $\mathcal{U}_k$}

\useasboundingbox
  ([xshift=-1.0cm]group c1r1.south west) rectangle
  ([yshift=0.5cm]group c4r1.north east);
\end{tikzpicture}
}
\noindent\vspace*{-0.2cm}%
    
    \caption{One iteration of Alg.~\ref{alg:main}. (a) $N$ points are sampled from $\hat{\mathcal{I}}(\theta_k)$. (b) The Poincaré map $\mathcal{P}$ partitions return points into those that leave the set ($\mathcal{V}_k \not\subset \hat{\mathcal{I}}(\theta_k)$) and those that return ($\mathcal{W}_k \subset \hat{\mathcal{I}}(\theta_k)$). (c) Their preimages form $\mathcal{Z}_k$ and $\mathcal{U}_k$, respectively. (d) $\hat{\mathcal{I}}(\theta_{k+1})$ is fit to $\mathcal{U}_k$. The algorithm repeats until the PAC bound satisfies a user-defined $\epsilon_{thresh}$.
    \vspace{-3mm}
    }
    \label{fig:hero}
\end{figure}

In contrast, we propose an approach to identify forward-invariant sets from sampled evaluations of the return map, providing finite-sample guarantees while accommodating any set representation.
This enables richer approximations of the invariant set geometry, yielding less conservative estimates. 
We evaluate sampled points from a candidate set under the return map and use the samples that satisfy forward invariance to refine the candidate set through a tractable optimization problem (see Fig.~\ref{fig:hero}).
The holdout method is then used to provide a risk estimate for the resulting prediction, based on an independent validation set.
This approach bounds the violation probability of the learned invariant set and ensures the bound holds with high confidence over the random sampling process.
Since our approach requires only closed-loop trajectory samples,
it can be readily parallelized using the same tools employed in large-scale robot controller training.
The main contributions of this work include:
\begin{itemize}
    \item We develop an iterative algorithm for computing invariant sets of the Poincar\'e return map of periodic hybrid systems, accommodating different set representations through a sampling-based optimization formulation of the discrete Poincar\'e-map dynamics.\vspace{-0.2cm}
    \item We provide probabilistic guarantees on finite-step invariance of the resulting sets and establish sufficient conditions for almost-sure termination of the algorithm in finite time.
    \vspace{-0.2cm}
    \item We demonstrate the proposed approach on three dynamical systems, including a Compass-Gait Walker model, and evaluate its performance against an existing baseline method.
\end{itemize}

\section{Preliminaries}
To contextualize our method, we first introduce periodic hybrid systems, followed by a brief discussion on the holdout method for deriving finite-sample guarantees.

\subsection{Hybrid Systems}
\label{sec:HybridSystems}
Consider a system with state $x \in \mathcal{X} \subseteq \mathbb{R}^n$. 
Let $h:\mathcal{X}\to\mathbb{R}$ be a continuously differentiable function defining the admissible domain $D = \{x \in \mathcal{X} \mid h(x) \ge 0\}$,
and the guard (or switching surface) $S = \{x \in \mathcal{X} \mid h(x) = 0,\ \dot{h}(x) < 0\}$.
The set $D$ is the region in which the continuous dynamics evolve and $S \subset D$ denotes the event surface where discrete transitions occur. The condition $\dot h(x) < 0$ ensures trajectories intersect the guard transversally. 

A reset map $\Delta:S \to D$ specifies the discrete transition from the pre-impact state $x^-$ to the post-impact state $x^+$. Let $\varphi_t(x_0)$ denote the flow of the continuous dynamics $\dot{x}=f(x)$ at time $t \ge 0$ with initial condition $x_0 \in D$. Hybrid trajectories are obtained by concatenating continuous evolution with discrete reset events.
Given a sequence of impact times $\{\tau_k\}_{k\ge1}$, with $x_k^-$, $x_k^+$ denoting the corresponding pre- and post-impact states, the hybrid trajectory satisfies $x_k^- = \varphi_{\tau_k-\tau_{k-1}}(x_{k-1}^+) \in S $ and $x_k^+ = \Delta(x_k^-)$,
for $k \ge 1$, with $\tau_0 = 0$ and initial condition $x_0 \in D$.
Between impacts, the trajectory evolves continuously according to $\varphi_{t-\tau_k}(x_k^+)$, $t \in [\tau_k,\tau_{k+1})$.

To analyze these systems, we use the guard surface $S$ as a natural Poincar\'e section.
Starting from a point $x \in S$, the reset map yields the post-impact state $x^+ = \Delta(x)$.
The system then evolves under the continuous dynamics until it intersects the guard $S$ again. Define the time-to-impact function $T(x) := \inf\{t > 0 \mid \varphi_t(\Delta(x)) \in S\}$.
The associated Poincar\'e return map $\mathcal{P}:S \rightarrow S$ is given by $\mathcal{P}(x) := \varphi_{T(x)}(\Delta(x))$.
This map induces a discrete-time dynamical system on the guard: 
\begin{align}
\label{eq:poincarereturnmap}
x_{k+1} = \mathcal{P}(x_k), \qquad x_k \in S.
\end{align}

A periodic orbit of the hybrid system corresponds to a fixed point of the Poincar\'e map, $x^\ast = \mathcal{P}(x^\ast)$, which
provides a reduced-order discrete-time representation of the hybrid dynamics on the section $S$.
Under the Poincar\'e map \eqref{eq:poincarereturnmap}, a set $\mathcal{I} \subset S$ is forward invariant if $x_0 \in \mathcal{I} \Rightarrow x_k \in \mathcal{I},  \forall k \in \mathbb{N}$.
In other words, trajectories initialized in $\mathcal{I}$ remain within the set at all future times.
We consider a weaker notion of invariance, termed \textit{finite-step or $n$-step invariance}, which requires $x_0 \in \mathcal{I} \Rightarrow x_n \in \mathcal{I}$ to only hold after $n$ iterations of the return map.
We focus primarily on one-step invariance and provide extensions to arbitrary $n$.

\subsection{Finite-Sample Guarantee}
\label{sec:holdout}
Sampling-based frameworks rely on a limited number of simulations, making it difficult to quantify confidence in resulting estimates.
Finite-sample guarantees address this challenge by providing rigorous statistical bounds for a fixed number of samples.
For black-box systems, \ac{pac} theory \citep{pacbound} provides a foundation for deriving such bounds.
We employ the holdout method, a general approach for establishing \ac{pac} guarantees under minimal assumptions.
The holdout method provides an a posteriori assessment 
of a solution by evaluating its accuracy on an independent test set \citep{langford2005tutorial}.
Specifically, a candidate solution is evaluated on $N$ i.i.d. samples, distinct from those used for training, and the number of observed violations $v$ is recorded.
Since the number of violations follows a binomial distribution, a binomial tail inversion can be used to bound the true violation probability at a user-defined confidence level $\beta$.
\begin{definition}[Binomial Tail Inversion]
\label{def:binomial_tail_inversion}
For $v$ violations out of $N$ scenarios and all $\beta \in (0, 1]$:
    \begin{equation}
        \overline{\text{Bin}}(v, N, \beta) = \max_{e}\Bigl\{ e : \text{Bin}\Bigl(v, N, e \Bigr) \geq \beta \Bigr\}, \,\,\mathrm{with} 
        \,\,\
        \text{Bin}\Bigl(v, N, e \Bigr) =  \sum_{j=0}^{v} \binom Nj e^j (1-e)^{N-j}.
        \label{eq:binominversion}
    \end{equation} 
    \label{def1}
\end{definition}
\vspace{-2mm}
In this paper, we aim to identify $\hat{\mathcal{I}}(\theta)$, an approximation of the invariant set satisfying the one-step invariance condition $\mathcal{P}(x) \in \hat{\mathcal{I}}(\theta), ~\forall x \in \hat{\mathcal{I}}(\theta)$.
For a fixed set $\hat{\mathcal{I}}(\theta)$ and $N$ i.i.d. samples drawn uniformly from $\hat{\mathcal{I}}(\theta)$, the holdout method yields a \ac{pac} guarantee in the form:
\begin{equation}
\label{eq:pacbound}
    \mathbf{P}^N(\mathbf{P}_{x \sim \mathrm{Uniform}(\hat{\mathcal{I}}(\theta))}(\mathcal{P}(x) \notin \hat{\mathcal{I}}(\theta))> \epsilon) \leq \beta,
\end{equation}
where $\epsilon = \overline{\text{Bin}}(v, N, \beta)$.
Importantly, this guarantee requires $\hat{\mathcal{I}}(\theta)$ to be fixed independently of the evaluation samples. We use \eqref{eq:pacbound} as a termination condition during invariant-set estimation and to validate the final invariant set (see Sec.~\ref{sec:algorithmdetails}).

\section{Finite-Step Invariant Set Identification}\label{sec:method}
Enforcing invariance is generally intractable without approximations or assumptions on the structure of $\mathcal{P}$.
Therefore, we aim to find invariant set estimates represented as level sets, $\hat{\mathcal{I}}(\theta) = \{x \in \mathbb{R}^{n_x} : g(x, \theta) \leq 0 \}$, where $g : \mathbb{R}^{n_x} \times \mathbb{R}^{n_\theta} \rightarrow \mathbb{R}$ defines the set representation and $\theta$ parameterizes the class of admissible set estimators. 
Ideally, we would formulate an optimization problem to maximize the volume of $\hat{\mathcal{I}}(\theta)$, subject to one-step forward invariance under the Poincar\'e return map $\mathcal{P}$:
\begin{equation}\label{eq:idealopt}
\max_{\theta} \ \text{Vol}(\hat{\mathcal{I}}(\theta)) \;\; \text{s.t.} \;\; \mathcal{P}(x) \in \hat{\mathcal{I}}(\theta),\ \forall x \in \hat{\mathcal{I}}(\theta).
\end{equation}
However, solving \eqref{eq:idealopt} is challenging because forward invariance must be verified over all states in $\hat{\mathcal{I}}(\theta)$, and volume maximization is generally nonconvex. 
Moreover, under black-box system assumptions, explicit gradient information is not available to facilitate convergence.
To address these limitations, we solve a relaxation of the optimization problem using a sampling-based algorithm that relies on stochastic exploration and function evaluations (i.e., simulation outputs under randomized inputs) to approximate optimal solutions in nonconvex domains. We require the resulting invariant set to satisfy the invariance condition with high probability, characterized by a PAC guarantee \eqref{eq:pacbound}.

We relax forward invariance 
to a set-containment problem, solving for a minimum-volume set $\hat{\mathcal{I}}(\theta)$ that contains all points satisfying one-step invariance.
This minimization strategy relies on the assumption that the initial set estimate at the start of the algorithm contains the true invariant set.
\begin{assumption}
\label{assump:initset}
    The initial set estimate contains
     the true invariant set, $\mathcal{I} \subset \hat{\mathcal{I}}(\theta_0)$.
\end{assumption}
Starting with this initial estimate $\hat{\mathcal{I}}(\theta_0)$, our iterative approach refines the invariant set by filtering input-output samples and refining the set through repeatedly solving an optimization problem (Sec. \ref{sec:algorithmdetails}).
Alg.~\ref{alg:main} summarizes the procedure, with further details provided in the following section. 


\begin{algorithm}[h!]
\caption{Finite-Step Invariant Set Identification}
\label{alg:main}
\begin{algorithmic}[1]
\REQUIRE Poincaré return map $\mathcal{P} : S \to S$,
         sample count $N$,
         initial set parameters $\theta_0$,
         confidence $ \beta \in (0,1)$, and 
         desired accuracy $1 - \beta^{1/N} \leq \epsilon_{thresh} < 1$
\ENSURE $\hat{\mathcal{I}}(\theta^*) \subseteq S$ such that, with confidence $1-\beta$, $\mathcal{P}(x) \in \hat{\mathcal{I}}(\theta^*)$ with probability at least $1-\epsilon^*$ for all $x \in \hat{\mathcal{I}}(\theta^*)$

\STATE Initialize $k \leftarrow 0, \epsilon_k=\infty$
\WHILE{\textbf{true}}
    \STATE \textbf{Step 1: Sample uniformly from current set estimate}
    \STATE $\quad$ Define $\hat{\mathcal{I}}(\theta_k) = \{x \in S : g(x, \theta_{k}) \leq 0 \}$ and draw $\{x_i\}_{i=1}^{N} \sim \operatorname{Uniform}(\hat{\mathcal{I}}(\theta_k))$
    \STATE \textbf{Step 2: Apply Poincaré return map}
    \STATE $\quad$ $x_i' \leftarrow \mathcal{P}(x_i), \quad \forall\, i \in \{1, \ldots, N\}$
    \STATE \textbf{Step 3: Partition by containment of return points}
    \STATE $\quad$ $\mathcal{V}_k \leftarrow \{ x_i' \mid x'_i \notin \hat{\mathcal{I}}(\theta_k)\}$, \hspace{0.5mm} \hspace{0.5mm} $\mathcal{U}_k \leftarrow \{ x_i \mid x'_i\in \hat{\mathcal{I}}(\theta_k)$\}
    \vspace{1mm}
    \STATE 
    $\quad \epsilon_k \leftarrow \overline{\text{Bin}}(|\mathcal{V}_k|, N, \beta)$
    \STATE $\quad$\textbf{if} $\epsilon_k \leq \epsilon_{thresh}$ \textbf{then}
        \STATE $\quad \quad k \leftarrow k+1$
        \STATE $\quad \quad$ \textbf{break}
    \STATE $\quad$\textbf{else}    
        \STATE $\quad$$\quad$\textbf{Step 4: Fit minimum-volume covering of $\mathcal{U}_k$}
        \vspace{1mm}
        \STATE $\quad$$\quad$
            $\begin{aligned}[t]
              \theta_{k+1} \leftarrow \arg\min_{\theta'}\ & \mathrm{Vol}(\theta') \;\; \text{s.t.} \;\;  g(x_i,\theta')\le 0,\ \forall x_i\in\mathcal{U}_k
            \end{aligned}$
    \STATE $\quad$\textbf{end if}
    \STATE $\quad k \leftarrow k+1$
\ENDWHILE
\STATE $\theta^* \leftarrow \theta_{k-1}$
\STATE \textbf{Step 5: Verification}
\STATE $\quad$Draw $\{x_i\}_{i=1}^{N} \sim \operatorname{Uniform}(\hat{\mathcal{I}}(\theta^*))$, $\quad$ $x'_i \leftarrow \mathcal{P}(x_i), \forall i \in \{1,...,N\}$
\STATE $\quad$$\mathcal{V}^*=\{x_i' \mid x'_i \notin \hat{\mathcal{I}}(\theta^*)\}$
\STATE $\quad \epsilon^* \leftarrow \overline{\text{Bin}}(|\mathcal{V}^*|, N, \beta)$
\RETURN $\hat{\mathcal{I}}(\theta^*) = \{x \in S : g(x, \theta^*) \leq 0 \}$, $\epsilon^*$
\end{algorithmic}
\end{algorithm}

\vspace{-3mm}
\subsection{Sampling-Based Algorithm Overview}
\label{sec:algorithmdetails}

\textbf{Generation of Input-Output Samples.} We draw $N$ i.i.d. samples $\{x_i\}_{i=1}^N$ from a uniform distribution defined over the current candidate invariant set $\hat{\mathcal{I}}(\theta_k)$
(Step 1, Alg. \ref{alg:main}).
Each sample is propagated through the Poincar\'e return map $\mathcal{P}$, yielding outputs $\{x'_i = \mathcal{P}(x_i)\}_{i=1}^{N}$ (Step 2, Alg. \ref{alg:main}).
We then classify each input-output pair based on their containment in $\hat{\mathcal{I}}(\theta_k)$, i.e., whether $g(x_i', \theta_{k}) \leq 0, \forall i=1,...,N$ (Step 3, Alg. \ref{alg:main}).
Pairs satisfying this condition are classified as \textit{inliers}, with their inputs 
collected into $\mathcal{U}_k$.
The remaining pairs are classified as \textit{outliers}, with their outputs collected in $\mathcal{V}_k$.
Since these operations are independent across samples, Steps 1-3 can be parallelized, reducing runtime when evaluating $\mathcal{P}$ is expensive.

\vspace{2mm} \noindent 
\textbf{Certification of Candidate Invariant Set.} We certify the candidate set at every iteration $k$ using a finite-sample guarantee obtained through a binomial tail inversion. 
At each iteration, we compute $\epsilon_k=\overline{\text{Bin}}(|\mathcal{V}_k|, N, \beta)$ for a user-defined confidence level $\beta$ (Line 9, Alg. \ref{alg:main}).
If $\epsilon_k$ satisfies the accuracy threshold $\epsilon_{thresh}$, the algorithm terminates and returns $\hat{\mathcal{I}}(\theta^*)=\hat{\mathcal{I}}(\theta_k)$ as the final invariant set.
However, utilizing a single set of samples for Steps 1-4 introduces a dependency between the returned set and the samples used for certification, potentially underestimating the error of $\hat{\mathcal{I}}(\theta^*)$.
We therefore use $\epsilon_k$ only as a stopping criterion and establish the PAC guarantee in Eq. \eqref{eq:pacbound} through an independent validation of the returned set (Step 5, Alg. \ref{alg:main}). Specifically, we evaluate $\hat{\mathcal{I}}(\theta^*)$ on a fresh set of $N$ i.i.d. input-output pairs $\{(x_i, x_i')\}_{i=1}^N$ where $x_i \sim \text{Uniform}(\hat{\mathcal{I}}(\theta^*))$.
This independent validation bounds the true error of the final invariant set, which may exceed $\epsilon_{thresh}$, since the stopping iteration was selected adaptively.

If $\epsilon_k$ does not satisfy the threshold $\epsilon_{thresh}$, the candidate set is refined using the sampling-based optimization procedure described below. 
Although Alg.~\ref{alg:main} is presented for identifying one-step forward-invariant sets with probabilistic guarantees, it readily extends to invariance over an arbitrary number of steps.
Specifically, the $n$-step guarantee bounds the probability that a test sample, initialized within the candidate set, lies outside the set after exactly $n$ successive applications of $\mathcal{P}$.
In Appendix \ref{app:verification}, we analyze how $\epsilon$ evolves with increasing $n$, providing insight into the robustness of the one-step guarantee.

\begin{remark}\label{corr:smallestepsilon}
    The smallest threshold $\epsilon_{thresh}$ for which Alg.~\ref{alg:main} can terminate is $\epsilon_{thresh} = 1 - \beta^{1/N}$, which follows directly from solving \eqref{eq:binominversion} for $\epsilon_k$ when $|\mathcal{V}_k| = 0$. This imposes constraints on $\epsilon_{thresh}$, as reflected in Alg.~\ref{alg:main}. The selection of $\epsilon_{thresh}$ is further discussed in Appendix \ref{appendix:discussion}.
\end{remark} 

\vspace{2mm}
\noindent 
\textbf{Refinement of Candidate Set.} If the candidate invariant set fails to satisfy the user-defined accuracy threshold, the algorithm
solves a sampling-based optimization problem over the new set of inliers $\mathcal{U}_k$.
Specifically, it computes the minimum-volume set enclosing all points in $\mathcal{U}_k$, 
minimizing a volume proxy of $\hat{\mathcal{I}}(\theta_{k+1})$
(Step 4, Alg. \ref{alg:main}).
By invariance, any sampled point that falls inside $\mathcal{I}$, remains in $\mathcal{I}$ under the Poincar\'e map, and is therefore
classified as an inlier by 
Alg.~\ref{alg:main}. 
Consequently, all sampled points of $\mathcal{I}$ are included in 
the updated invariant set estimate $\hat{\mathcal{I}}(\theta_{k+1})$. 

\vspace{-2mm}
\subsection{Exploring Set Representations}
Alg.~\ref{alg:main} provides a general procedure for constructing invariant sets, independent of the set representation $g(x, \theta)$. The choice of representation depends on the underlying system and the desired balance between approximation accuracy and computational efficiency.
Here, we introduce two representations of increasing expressiveness: ellipsoids and radial basis functions (RBFs).
\footnote{We provide our implementation at: \url{https://github.com/dynamicmobility/finite-step-invariance}}

We first consider ellipsoids, which provide compact representations and admit tractable optimization formulations. Specifically, we parameterize the ellipsoid using a positive definite matrix $A \in \mathbb{R}^{n \times n}$ and vector $b \in \mathbb{R}^n$, such that $\hat{\mathcal{I}}(\theta) = \mathcal{E}(\theta) = \{x \in \mathbb{R}^n \mid ||Ax-b||_2 \leq 1\}$, and minimize $- \log \det(A)$ as a convex volume proxy in Line 15 of Alg.~\ref{alg:main}. This parameterization preserves convexity of the resulting optimization problem, improving numerical stability and solver efficiency.
However, ellipsoids may poorly approximate nonconvex sets, enclosing substantial regions outside the true invariant set and resulting in low-accuracy estimates. 

Therefore, we alternatively represent the invariant set using a finite collection of radial basis functions (RBFs). Each RBF is defined by a Gaussian function parameterized by its center $\mu$, width $\sigma$, and level-set threshold $\gamma$, yielding the following optimization problem for $m$ RBFs, replacing Line 15 in Alg. \ref{alg:main}:
\begin{equation}
\label{eq:rbfs_opt}
 \underset{\mu \in \mathbb{R}^{m \times n_{x}},\; \sigma \in (0,\infty)^m}{\min} \hspace*{1mm} 
\sum^m_{i=1} \sigma_i^2\\
\hspace*{4mm} \text{s.t. }
\sum^m_{i=1} e^{-\frac{1}{2}\frac{(x^{(j)}- \mu_i)^2}{\sigma_i^2}}-\gamma \geq 0, \forall j=1,\dotsc,|\mathcal{U}|.
\end{equation}
By summing $m$ RBFs, their overlapping tails represent geometries expressive beyond the union of ellipsoids. 
This enables tighter approximations of nonconvex invariant sets.
However, RBFs require solving a nonconvex optimization problem, which reduces convergence stability and increases computational cost. 
We present two examples that highlight the tradeoffs of each set representation. 

\textbf{Convex Expander-Contractor (CEC).}
First, we present 
a 2-dimensional Convex Expander-Contractor (CEC) with the following dynamics:
\vspace{-1mm}
\begin{gather}
    x_{k+1} = \mathcal{P}(x_k) = (x_k-c)\sqrt{(x_k-c)^TM(x_k-c)} + c.
\end{gather}
\vspace{-1mm}
This system scales vectors relative to the fixed point $c$ by its $M$-weighted norm, where $M \succ 0$ is positive definite.
Points inside the unit $M$-norm ball centered at $c$ contract towards $c$, while points outside the ball expand.
Consequently, the ellipsoid $\{x:(x-c)^TM(x-c)\leq 1 \}$ is both an invariant set and a region of attraction for $c$.
In all simulations, we set \mbox{$c=[1, 1]^T$} and $M=\begin{psmallmatrix}2 & 1\\ 1 & 1\end{psmallmatrix}$.
We first initialize Alg. \ref{alg:main} with an ellipsoid parameterized by $b_0=[0, 0]^T$ and $A_0 = 0.1\begin{psmallmatrix}\sqrt{10} & 0\\ 0 & \sqrt{10}\end{psmallmatrix}$,
confidence level $\beta=10^{-9}$, sample count $N=2000$, and accuracy threshold $\epsilon_{thresh}=1-\beta^{1/N}=1.031\times10^{-2}$.
On average, across 10 trials,
the algorithm recovers an invariant ellipsoid with $\epsilon^*=0.0111$, in 7.5 iterations.
Additionally, when run with Eq. \eqref{eq:rbfs_opt} and $\epsilon_{thresh}=0.06$, Alg. \ref{alg:main} recovers 3 RBFs, with an average $\epsilon^*=0.054$ after 32.2 iterations over 10 trials.
Fig. \ref{fig:cecsolution} in Appendix \ref{app:results}
presents the recovered invariant sets.

\begin{figure}[t]
    \centering
    \resizebox{\linewidth}{!}{
\usetikzlibrary{calc}
\pgfplotsset{compat=1.18}
\newcommand{\figdir}{l4dc_figures/pdfs/}

\pgfplotsset{
  colormap={plasma}{
    rgb=(0.050383,0.029803,0.527975)
    rgb=(0.254627,0.013882,0.615419)
    rgb=(0.417642,0.000564,0.658390)
    rgb=(0.562738,0.051545,0.641509)
    rgb=(0.692840,0.165141,0.564522)
    rgb=(0.798216,0.280197,0.469538)
    rgb=(0.881443,0.392529,0.383229)
    rgb=(0.949217,0.517763,0.295662)
    rgb=(0.987621,0.652025,0.211364)
    rgb=(0.992440,0.809579,0.145357)
    rgb=(0.940015,0.975158,0.131326)
  },
}

\definecolor{cEll}{rgb}{0.2546,0.0139,0.6154}   
\definecolor{cRBF}{HTML}{FF8C00}                
\definecolor{cROA}{HTML}{FBDDE6}                

\pgfplotsset{
  panel/.style={
    width=\pw, height=\pw,            
    axis on top,
    scale only axis,
    tick align=outside,
    tick style={black, line width=0.4pt},
    every tick label/.append style={font=\scriptsize},
    label style={font=\footnotesize},
    xlabel style={yshift=3pt},
    ylabel style={yshift=-7pt},
    xtick pos=left, ytick pos=left,
    xticklabel style={/pgf/number format/.cd, fixed, fixed zerofill, precision=1},
    yticklabel style={/pgf/number format/.cd, fixed, fixed zerofill, precision=1},
    colormap name=plasma,
  },
  itbar/.style={
    colorbar horizontal,
    colorbar style={
      at={(0.5,1.04)}, anchor=south,       
      height=0.2cm,
      xtick pos=upper, xticklabel pos=upper,
      xticklabel style={font=\scriptsize, /pgf/number format/.cd, fixed, precision=0},
      title={Iteration}, title style={font=\footnotesize, yshift=0.38cm},
      tick align=outside, tick style={black, line width=0.4pt},
    },
  },
}

\newcommand{\itminA}{0}   \newcommand{\itmaxA}{150}   
\newcommand{\itminB}{0}   \newcommand{\itmaxB}{10}    

\newcommand{\pa}{0.95cm}
\newcommand{\gap}{1.25cm}   

\pgfmathsetlengthmacro{\pw}{(\linewidth - 2*\pa - 2*\gap)/3}
\pgfmathsetlengthmacro{\pitch}{\pw + \gap}

\newcommand{\figshift}{1cm}
\begingroup\noindent\hspace*{-\figshift}%
\begin{tikzpicture}
    
\useasboundingbox (0,-1.55cm) rectangle (\linewidth, \pw+1.25cm);

\begin{axis}[
  panel, itbar, point meta min=\itminA, point meta max=\itmaxA,
  at={(\pa,0)}, anchor=south west, name=p1,
   xmin=-3.9992, xmax=4.3420, ymin=-4.2355, ymax=4.1057,
  xtick={-3,0,3}, ytick={-3,0,3},
  xlabel={$x_1$}, ylabel={$x_2$}
]
  \addplot graphics[ xmin=-2.667055, xmax=2.574026, ymin=-2.552118, ymax=2.688963]
    {\figdir DogBone_ellipsoid_progression_bare.pdf};
\end{axis}
\node[anchor=north, font=\small] at ([yshift=-0.95cm]p1.south) {(a)};

\begin{axis}[
  panel, itbar, point meta min=\itminB, point meta max=\itmaxB,
  at={(\pa+\pitch,0)}, anchor=south west, name=p2,
  xmin=-3.9992, xmax=4.3420, ymin=-4.2355, ymax=4.1057,
  xtick={-3,0,3}, ytick={-3,0,3},
  xlabel={$x_1$}, ylabel={$x_2$}
]
  \addplot graphics[xmin=-3.9992, xmax=4.3420, ymin=-4.2355, ymax=4.1057]
    {\figdir DogBone_rbf_progression_bare.pdf};
\end{axis}
\node[anchor=north, font=\small] at ([yshift=-0.95cm]p2.south) {(b)};

\begin{axis}[
  panel, at={(\pa+2*\pitch,0)}, anchor=south west, name=p3,
  xmin=-0.9, xmax=0.9, ymin=-0.9, ymax=0.9,
  xtick={-0.8,0,0.8}, ytick={-0.8,0,0.8},
  xlabel={$x_1$}, ylabel={$x_2$}
]
  \addplot graphics[xmin=-0.9, xmax=0.9, ymin=-0.9, ymax=0.9]
    {\figdir DogBone_comparison_bare.pdf};
\end{axis}
\node[anchor=north, font=\small] at ([yshift=-0.95cm]p3.south) {(c)};

\pgfmathsetlengthmacro{\lgy}{0.04*\pw}     
\pgfmathsetlengthmacro{\lgw}{\pw/3}         
\begin{scope}[shift={(\pa+2*\pitch, \pw+\lgy)}, font=\scriptsize]   
  \fill[cEll] (0,0)         rectangle (\lgw,0.2cm);
  \fill[cRBF] (\lgw,0)       rectangle (2*\lgw,0.2cm);
  \fill[cROA] (2*\lgw,0)     rectangle (3*\lgw,0.2cm);
  \draw[black, line width=0.4pt] (0,0) rectangle (3*\lgw,0.2cm);
  \foreach \i/\lab in {0/Ellipsoid,1/RBF,2/True set}{
    \draw[black, line width=0.4pt] ({(\i+0.5)*\lgw},0.2cm) -- ++(0,0.1cm);
    \node[anchor=base, black, inner sep=0pt] at ({(\i+0.5)*\lgw},0.38cm) {\lab};  
  }
  \node[anchor=south, black, font=\footnotesize] at ({1.5*\lgw},0.80cm) {Final sets};
\end{scope}

\end{tikzpicture}\par\endgroup
    }
    \caption{Invariant sets of Nonconvex Expander-Contractor (NEC) system. (a) Sequence of ellipsoid iterates $\hat{\mathcal{E}}(\theta_0), \hat{\mathcal{E}}(\theta_1), \ldots, \hat{\mathcal{E}}(\theta^*)$. (b) Sequence of RBF iterates $\hat{\mathcal{R}}(\theta_0), \hat{\mathcal{R}}(\theta_1), \ldots, \hat{\mathcal{R}}(\theta^*)$. (c) The true invariant set (light pink), consisting of two circles, is closely approximated by the RBF representation (orange dashed). The ellipsoid (dark purple) results in an under-approximation with lower accuracy.
    \vspace{-3mm}
    }
    \label{fig:nesolution}
\end{figure}

\textbf{Nonconvex Expander-Contractor (NEC).} We next introduce the Nonconvex Expander-Contractor (NEC), whose invariant set consists of two regions connected by a narrow bridge, which cannot be captured by an ellipsoid.
The NEC dynamics are defined by the following piecewise function:
\begin{gather*}
    \mathcal{P}(x) = 
    \begin{cases}
\frac{1}{2}(x+c_1) & \textrm{if}~\|x-c_1\|_2<r,\\
\frac{1}{2}(x+c_2) & \textrm{if}~\|x-c_2\|_2<r,\\
\kappa x & \textrm{otherwise},
\end{cases}
\end{gather*}
for $x, c_1, c_2 \in \mathbb{R}^2$ and $r > 0, \kappa>1$.
The invariant set is a pair of
circles of radius $r$, centered at $c_1$ and $c_2$,
with $c_1=[-0.6, 0]^T, c_2=[0.6, 0]^T, r=0.6$, and $\kappa=1.3$.
To compute a forward-invariant ellipsoid, we set $\epsilon_{thresh}=0.07$, $\beta=10^{-9}$, and $N=2000$,
and provide an initial guess by fitting an ellipse to a uniform sampling of 100 points over the interval $[-2, 2]^2$.
On average, across 10 trials the algorithm recovers an ellipsoid with $\epsilon^*=0.077$, in 138.6 iterations.
This system highlights the limitations of ellipsoidal representations for nonconvex invariant sets.
When run with two RBFs and $\epsilon_{thresh}=0.04$, Alg. \ref{alg:main}
nearly recovers the true NEC invariant set 
reliably across 10 trials with $\epsilon^* = 0.0406$ in 17.7 iterations.
Fig. \ref{fig:nesolution} compares the resulting set estimates for both representations.

\section{Application to Compass-Gait Walker}
\label{sec:compass_gait}
We demonstrate our algorithm on the identification of discrete-time invariant sets of the Poincar\'e map for the Compass-Gait Walker. 
Identifying invariant sets for legged systems is used to characterize the operating conditions under which legged robots can sustain periodic walking.
The Compass-Gait Walker is a 2 degree-of-freedom model that captures the hybrid dynamics of walking.
At each step, the stance leg is pinned to the ground, while the swing leg rotates about the hip.
Upon ground impact, the legs exchange roles through an inelastic collision.
When walking downhill, the Compass-Gait Walker exhibits a passively stable limit cycle, 
in which energy lost during each step and ground impact is exactly counteracted by the potential energy gained from going downhill.
Since the guard $S$ is defined at impact, this periodic gait corresponds to a stable fixed point $x^\ast \in S$ of the Poincar\'e map, whose local stability can be analyzed through linearization.
Specifically, quadratic Lyapunov functions of the linearized map provide ellipsoidal sublevel sets that approximate its region of attraction.
This motivates the use of ellipsoids as a set representation.
Despite extensive study of the Compass-Gait Walker \citep{goswami1998compass, manchester2010regionsattractionhybridlimit, choi2022regions}, the system's Poincar\'e return map has no closed-form expression, which makes identifying its true invariant set difficult.
Thus, characterizing the invariant set requires numerical simulation. 
We demonstrate our approach to recover a probabilistic finite-step ellipsoidal invariant set using only forward simulations.
\begin{figure}[t]
    \centering
    \noindent\hspace*{-1cm}
    \resizebox{\linewidth}{!}{
        \pgfplotsset{compat=1.18}

\newcommand{\figdir}{l4dc_figures/pdfs/}
\newcommand{\pw}{4cm}   
\newcommand{\ph}{4cm}   

\definecolor{cU}{rgb}{0.692840,0.165141,0.564522}   
\definecolor{cZ}{rgb}{0.915330,0.455146,0.339446}   
\definecolor{cW}{rgb}{0.254627,0.013882,0.615419}   
\definecolor{cV}{rgb}{0.990031,0.730802,0.178361}   

\newcommand{\legline}[1]{\tikz[baseline=-0.5ex]{\draw[#1] (0,0)--(0.7,0);}}
\newcommand{\legdot}[1]{\tikz[baseline=-0.5ex]{\fill[#1] (0,0) circle (2.6pt);}}

\pgfplotsset{
  colormap={plasma}{
    rgb=(0.050383,0.029803,0.527975)
    rgb=(0.254627,0.013882,0.615419)
    rgb=(0.417642,0.000564,0.658390)
    rgb=(0.562738,0.051545,0.641509)
    rgb=(0.692840,0.165141,0.564522)
    rgb=(0.798216,0.280197,0.469538)
    rgb=(0.881443,0.392529,0.383229)
    rgb=(0.949217,0.517763,0.295662)
    rgb=(0.987621,0.652025,0.211364)
    rgb=(0.992440,0.809579,0.145357)
    rgb=(0.940015,0.975158,0.131326)
  },
}

\pgfplotsset{
  panel/.style={
    width=\pw, height=\ph,
    axis on top,
    scale only axis,
    tick align=outside,
    tick style={black},
    every tick label/.append style={font=\normalsize},
    label style={font=\Large},
    xlabel style={yshift=2pt},
    legend style={font=\small, fill=white, fill opacity=0.9, text opacity=1,
                  draw=black!30, cells={anchor=west}},
    legend pos=north east,
    colormap name=plasma,
  },
  zf3/.style={/pgf/number format/.cd, fixed, fixed zerofill, precision=3},
  zf0/.style={/pgf/number format/.cd, fixed, precision=0},
}

\resizebox{\linewidth}{!}{%
\begin{tikzpicture}
\begin{scope}[xshift=-3cm]
\node[anchor=south, draw=black!30, fill=white, inner sep=4pt, font=\Large]
  at (14.9cm, 4.7cm) {%
  \legline{black, dashed, thick}\;$\mathcal{E}_0$\quad
  \legline{black, thick}\;$\mathcal{E}^*$\quad
  \legdot{cU}\;$\mathcal{U}$\quad
  \legdot{cZ}\;$\mathcal{Z}$\quad
  \legdot{cW}\;$\mathcal{W}$\quad
  \legdot{cV}\;$\mathcal{V}$%
};

\begin{axis}[
  panel, at={(7.4cm,0)}, anchor=south west,
  xmin=-0.04352, xmax=0.04352, ymin=-1.085, ymax=1.085,
  xtick pos=left,
  ytick pos=left,
  xtick={-0.03,0,0.03},
  ytick={-1.0, 0, 1.0},
  scaled x ticks=false,
  ylabel style={yshift=-10pt}, 
  scaled y ticks=false,
  xticklabel style={/pgf/number format/.cd, fixed, fixed zerofill, precision=2},
  yticklabel style={/pgf/number format/.cd, fixed, fixed zerofill, precision=1},
  xlabel={$x_1$}, ylabel={$x_2$},
]
  \addplot graphics[xmin=-0.04352, xmax=0.04352, ymin=-1.085, ymax=1.085]
    {\figdir Systems.CompassGaitWalker_ellipsoid_log_20261112_195318_614395_seed0_invariant_combined_bare_x1x2.pdf};
\end{axis}

\begin{axis}[
  panel, at={(12.9cm,0)}, anchor=south west, name=p3,
  xmin=-0.04352, xmax=0.04352, ymin=-0.141, ymax=0.1411,
  xtick pos=left,
  ytick pos=left,
  xtick={-0.03,0,0.03}, xticklabel style={zf3},
  ytick={-0.1,0,0.1},
  yticklabel style={zf3},
  ylabel style={yshift=-10pt}, 
  scaled x ticks=false,
  scaled y ticks=false,
  xticklabel style={/pgf/number format/.cd, fixed, fixed zerofill, precision=2},
  yticklabel style={/pgf/number format/.cd, fixed, fixed zerofill, precision=1},
  xlabel={$x_1$}, ylabel={$x_3$},
]
  \addplot graphics[xmin=-0.04352, xmax=0.04352, ymin=-0.141, ymax=0.1411]
    {\figdir Systems.CompassGaitWalker_ellipsoid_log_20261112_195318_614395_seed0_invariant_combined_bare_x1x3.pdf};
\end{axis}
\begin{axis}[
  panel, at={(18.35cm,0)}, anchor=south west,
  xmin=-1.085, xmax=1.085, ymin=-0.141, ymax=0.1411,
  xtick={-1.0,0,1.0}, xticklabel style={zf0},
  xtick pos=left,
  ytick pos=left,
  ylabel style={yshift=-10pt}, 
  ytick={-0.1, 0, 0.1},
  yticklabel style={zf3},
  scaled x ticks=false,
  scaled y ticks=false,
  xticklabel style={/pgf/number format/.cd, fixed, fixed zerofill, precision=2},
  yticklabel style={/pgf/number format/.cd, fixed, fixed zerofill, precision=1},
  xlabel={$x_2$}, ylabel={$x_3$},
  colorbar,
  point meta min=1, point meta max=49,
  colorbar style={ytick={10,20,30,40,50}, ylabel={Iteration},
                  width=0.25cm, yticklabel style={font=\small},
                  ylabel style={font=\small}},
]
  \addplot graphics[xmin=-1.085, xmax=1.085, ymin=-0.141, ymax=0.1411]
    {\figdir Systems.CompassGaitWalker_ellipsoid_log_20261112_195318_614395_seed0_invariant_combined_bare_x2x3.pdf};
\end{axis}

\end{scope}
\end{tikzpicture}%

}
\vspace{-1em}
        }
    \caption{Sampling-based, finite-step invariant set computation for the Compass-Gait Walker. The sequence of ellipsoid iterates $\hat{\mathcal{E}}(\theta_0), \hat{\mathcal{E}}(\theta_1), \ldots, \hat{\mathcal{E}}(\theta^*)$ and verification of the final set $\hat{\mathcal{E}}(\theta^*)$. 
    With accuracy threshold $\epsilon_{thresh} = 0.04$, 
    Alg. \ref{alg:main} converged in 49 iterations to an ellipsoid with volume $1.4\times10^{-3}$.
    \vspace{-3mm}
    }
    \label{fig:compass_gait}
\end{figure}

We exploit the Compass-Gait Walker's stable periodic orbit to initialize Alg. \ref{alg:main}.
We compute the Jacobian of the Poincar\'e map $J=D\mathcal{P}(x)$ around the stable fixed point $x^*$, using the automatic differentiation capability of \texttt{jax} \citep{jax2018github}.
The eigenvalues of $J$ (i.e., Floquet multipliers) characterize the local stability of $x^*$, which is asymptotically
stable if and only if $|\lambda_i| < 1$ for all $i$.
When this condition holds, the linearization of $\mathcal{P}$ can be used to compute a quadratic Lyapunov function for the linearized system,
using an approach inspired by discrete-time contraction theory \citep[Eq. 2.36]{bullo2022contraction}.
Since quadratic Lyapunov function sublevel sets are forward-invariant ellipsoids, this provides an initial estimate of the invariant set's shape under the true dynamics.

We solve a semi-definite program to generate a shaping matrix, which decays according to a minimum contraction rate based on the spectral radius of $J$, denoted by $b$:
\begin{align}\label{eq:initial_set_SDP}
    \min_{\mathbf{X}\succ \mathbf{I}} & \quad \operatorname{trace}(\mathbf{X}) \quad  \text{s.t. } \quad J^T \mathbf{X} J \preceq b^2 \mathbf{X}.
\end{align}
The matrix $\mathbf{X}$ characterizes the local contraction behavior of the dynamics around the fixed point. An ellipsoid defined by $\mathbf{X}$ contracts at rate $b$ under the linearized dynamics $J$.
We use this contraction geometry to construct an initial ellipsoid approximation, centered at $x^*$ and scaled by a user-defined factor $r$,
to ensure satisfaction of Assumption \ref{assump:initset}.
Given this initial set, over 10 trials, Alg. \ref{alg:main} converged on average in 63.8 iterations to an ellipsoid with error $\epsilon^*=0.0457$.
One such ellipsoid is illustrated in Fig. \ref{fig:compass_gait} as two-dimensional slices.

Additionally, we compare our proposed approach against the normotope-based method from \citep{madabushi2026}, which computes a reachable normotope tube around the limit cycle, converts its intersections with the guard surface into a collection of sliced normotopes, and formally certifies invariance by verifying that all reset slices map into a subset of the initial normotope.
As shown in Fig.~\ref{fig:cgw_comparison}, this approach yields an invariant set that is 10 times smaller than the one obtained by our proposed sampling-based method.
However, the sampling-based method yields an empirical error of 1.58\%, estimated from a Monte Carlo experiment with 13,000 samples, while the normotope-based method exhibits 0\% error.
Note that for this particular system, the discretization of the normotope tube results in 13 sliced normotopes, so the Monte Carlo experiment requires that the sample count is divisible by 13. 
Additional details on this comparison are provided in Appendix~\ref{app:results}.
\begin{figure}[t]
    \centering
    \hspace*{-3cm}
    \resizebox{\linewidth}{!}{
        \pgfplotsset{compat=1.18}

\newcommand{\figdir}{l4dc_figures/pdfs/}
\newcommand{\pw}{4.5cm}   
\newcommand{\ph}{4.5cm}   

\definecolor{cU}{rgb}{0.692840,0.165141,0.564522}   
\definecolor{cZ}{rgb}{0.915330,0.455146,0.339446}   
\definecolor{cW}{rgb}{0.254627,0.013882,0.615419}   
\definecolor{cV}{rgb}{0.990031,0.730802,0.178361}   

\newcommand{\legline}[1]{\tikz[baseline=-0.5ex]{\draw[#1] (0,0)--(0.7,0);}}
\newcommand{\legdot}[1]{\tikz[baseline=-0.5ex]{\fill[#1] (0,0) circle (2.6pt);}}

\pgfplotsset{
  colormap={plasma}{
    rgb=(0.050383,0.029803,0.527975)
    rgb=(0.254627,0.013882,0.615419)
    rgb=(0.417642,0.000564,0.658390)
    rgb=(0.562738,0.051545,0.641509)
    rgb=(0.692840,0.165141,0.564522)
    rgb=(0.798216,0.280197,0.469538)
    rgb=(0.881443,0.392529,0.383229)
    rgb=(0.949217,0.517763,0.295662)
    rgb=(0.987621,0.652025,0.211364)
    rgb=(0.992440,0.809579,0.145357)
    rgb=(0.940015,0.975158,0.131326)
  },
}

\pgfplotsset{
  panel/.style={
    width=\pw, height=\ph,
    axis on top,
    scale only axis,
    tick align=outside,
    tick style={black},
    every tick label/.append style={font=\normalsize},
    label style={font=\Large},
    xlabel style={yshift=2pt},
    legend style={font=\small, fill=white, fill opacity=0.9, text opacity=1,
                  draw=black!30, cells={anchor=west}},
    legend pos=north east,
    colormap name=plasma,
  },
  zf3/.style={/pgf/number format/.cd, fixed, fixed zerofill, precision=3},
  zf0/.style={/pgf/number format/.cd, fixed, precision=0},
}

\resizebox{\linewidth}{!}{%
\begin{tikzpicture}
\begin{axis}[
  panel, name=mainplot, at={(5cm,0)}, anchor=south west,
  xmin=-0.0271, xmax=0.0212, ymin=-0.889, ymax=0.6401,
  xtick pos=left,
  ytick pos=left,
  xtick={-0.02,0,0.02},
  ytick={-0.5, 0, 0.5},
  scaled x ticks=false,
  ylabel style={yshift=-8pt},
  scaled y ticks=false,
  xticklabel style={/pgf/number format/.cd, fixed, fixed zerofill, precision=2},
  yticklabel style={/pgf/number format/.cd, fixed, fixed zerofill, precision=1},
  xlabel={$x_1$}, ylabel={$x_2$},
]
  \addplot graphics[xmin=-0.0271, xmax=0.0212, ymin=-0.889, ymax=0.6401]
    {\figdir immrax_guard_projection_bare_z1z2.pdf};
\end{axis}

\begin{axis}[
  at={(5.3cm,4.3cm)}, anchor=north west,
  anchor=north west,
  scale only axis,
  xtick pos=left,
  ytick pos=right,
  width=1cm, height=1cm,
  axis on top,
  xmin=-0.00243, xmax=0.002633, ymin=-0.09668, ymax=0.09801,
  xtick={-0.002,0.002}, ytick={-0.05,0.05},
  scaled x ticks=false, scaled y ticks=false,
  xticklabel style={/pgf/number format/.cd, fixed, precision=3},
  yticklabel style={/pgf/number format/.cd, fixed, precision=2},
  tick label style={font=\tiny},
  tick align=outside,
  axis background/.style={fill=white, fill opacity=1},
  axis line style={draw=black, line width=0.4pt},
]
  \addplot graphics[xmin=-0.00243, xmax=0.002633, ymin=-0.09668, ymax=0.09801]
    {\figdir immrax_guard_projection_bare_zoom_z1z2.pdf};
\end{axis}
\hspace{0.75cm}

\begin{axis}[
  panel, at={(10.9cm,0)}, anchor=south west, name=p3,
  xmin=-0.0271, xmax=0.0212, ymin=-0.05534, ymax=0.06821,
  xtick pos=left,
  ytick pos=left,
  xtick={-0.02,0,0.02}, xticklabel style={zf3},
  ytick={-0.05,0,0.05},
  yticklabel style={zf3},
  ylabel style={yshift=-8pt}, 
  scaled x ticks=false,
  scaled y ticks=false,
  xticklabel style={/pgf/number format/.cd, fixed, fixed zerofill, precision=2},
  yticklabel style={/pgf/number format/.cd, fixed, fixed zerofill, precision=1},
  xlabel={$x_1$}, ylabel={$x_3$},
]
  \addplot graphics[xmin=-0.0271, xmax=0.0212, ymin=-0.05534, ymax=0.06821]
    {\figdir immrax_guard_projection_bare_z1z3.pdf};
\end{axis}


\begin{axis}[
  at={(15.1cm,4.3cm)}, anchor=north east,
  scale only axis,
  xtick pos=left,
  ytick pos=left,
  width=1cm, height=1cm,
  axis on top,
  xmin=-0.002386, xmax=0.002583, ymin=-0.008128, ymax=0.008142,
  xtick={-0.002,0.002}, ytick={-0.008,0.008},
  scaled x ticks=false, scaled y ticks=false,
  xticklabel style={/pgf/number format/.cd, fixed, precision=3},
  yticklabel style={/pgf/number format/.cd, fixed, precision=2},
  tick label style={font=\tiny},
  tick align=outside,
  axis background/.style={fill=white, fill opacity=1},
  axis line style={draw=black, line width=0.4pt},
]
  \addplot graphics[xmin=-0.002386, xmax=0.002583, ymin=-0.008128, ymax=0.008142]
    {\figdir immrax_guard_projection_bare_zoom_z1z3.pdf};
\end{axis}
\hspace{0.5cm}

\hspace{1cm}
\begin{axis}[
  panel, at={(16.35cm,0)}, anchor=south west,
  xmin=-0.889, xmax=0.6401, ymin=-0.05534, ymax=0.06821,
  xtick={-0.5,0,0.5}, xticklabel style={zf0},
  xtick pos=left,
  ytick pos=left,
  ylabel style={yshift=-8pt}, 
  ytick={-0.05, 0, 0.05},
  yticklabel style={zf3},
  scaled x ticks=false,
  scaled y ticks=false,
  xticklabel style={/pgf/number format/.cd, fixed, fixed zerofill, precision=2},
  yticklabel style={/pgf/number format/.cd, fixed, fixed zerofill, precision=1},
  xlabel={$x_2$}, ylabel={$x_3$},
]
  \addplot graphics[xmin=-0.889, xmax=0.6401, ymin=-0.05534, ymax=0.06821]
    {\figdir immrax_guard_projection_bare_z2z3.pdf};
\end{axis}


\begin{axis}[
  at={(20.6cm,4.3cm)}, anchor=north east,
  scale only axis,
  xtick pos=left,
  ytick pos=left,
  width=1cm, height=1cm,
  axis on top,
  xmin=-0.09668, xmax=0.098, ymin=-0.008128, ymax=0.008142,
  xtick={-0.09,0.09}, ytick={-0.008,0.008},
  scaled x ticks=false, scaled y ticks=false,
  xticklabel style={/pgf/number format/.cd, fixed, precision=3},
  yticklabel style={/pgf/number format/.cd, fixed, precision=2},
  tick label style={font=\tiny},
  tick align=outside,
  axis background/.style={fill=white, fill opacity=1},
  axis line style={draw=black, line width=0.4pt},
]
  \addplot graphics[xmin=-0.09668, xmax=0.098, ymin=-0.008128, ymax=0.008142]
    {\figdir immrax_guard_projection_bare_zoom_z2z3.pdf};
\end{axis}

\end{tikzpicture}%

}
\vspace{-1em}
        }
    \caption{Comparison of invariant sets recovered by the proposed method (gray ellipsoid) and the normotope-based method of \citep{madabushi2026} (plasma ellipsoids). The proposed approach recovers an invariant set 10 times larger, yielding a less conservative estimate at the cost of increased error, exhibiting an empirical escape rate of $1.58\%$ compared to $0\%$ for the normotope-based method of \citep{madabushi2026}, given 13,000 Monte Carlo samples.
    \vspace{-3mm}
    }
    \label{fig:cgw_comparison}
\end{figure}

\section{Termination Analysis}\label{sec:finitestepconvergence}
As demonstrated in Secs.~\ref{sec:method} and \ref{sec:compass_gait}, the convergence speed of Alg.~\ref{alg:main} depends on how well the chosen set representation captures the true invariant set and on the prescribed accuracy threshold. In this section, we analyze the algorithm's convergence and establish conditions under which termination can be expected in practice.
To this end, we reason about the true rate of escape of $\hat{\mathcal{I}}(\theta_k)$, which is the probability that a uniformly sampled point leaves the candidate set after one application of $\mathcal{P}$,
\begin{align}\label{eq:trueescaperate}
    r(\hat{\mathcal{I}}(\theta_k)) = \mathbf{P}_{x \sim \mathrm{Uniform}(\hat{\mathcal{I}}(\theta_k))}(\mathcal{P}(x) \notin \hat{\mathcal{I}}(\theta_k))  = 1 - \frac{\mu(\{x \in \hat{\mathcal{I}}(\theta_k) : \mathcal{P}(x) \in \hat{\mathcal{I}}(\theta_k) \})}{\mu(\hat{\mathcal{I}}(\theta_k))},
\end{align}
where $\mu(\cdot)$ is the Lebesgue measure on $S \subseteq \mathbb{R}^n$ (e.g., volume). We assume $\mathcal{P}$ is measurable and $\mu{(\hat{\mathcal{I}}(\theta_k))}>0$.
Termination of Alg.~\ref{alg:main} requires the existence of a constant $r_{\max} < 1$ such that, at every iteration $k$, $r(\hat{\mathcal{I}}(\theta_k)) \leq r_{\max}<1$.
Intuitively, if a draw yields too few returning points to fit a set in Step 4 of Alg.~\ref{alg:main}, the algorithm resamples. Since $r(\hat{\mathcal{I}}(\theta_k))<1$, each draw succeeds with probability bounded away from 0 independently of $k$, and Step 4 eventually succeeds.
Although this assumption is restrictive, its practical implications can be partially mitigated through increased sampling, which significantly improves the likelihood of sampling points within the true invariant set. In all presented examples, uniformly sampling with $N=2000$ satisfied this assumption.

Alg.~\ref{alg:main} terminates when the certified bound on the candidate invariant set's escape rate satisfies $\epsilon_k \leq \epsilon_{thresh}$.
Since each iteration utilizes fresh samples, termination at any given iteration is a random event.
Therefore, we model the number of iterations as a random variable and show that it is finite with probability $1$.
Let $\hat{\mathcal{I}}(\theta_k)$ denote the candidate invariant set at iteration $k+1$, with certified bound $\epsilon_{k}$, and
define the termination iteration as $T = \min \{k \geq 1 : \epsilon_{k-1} \leq \epsilon_{thresh} \}$.
Conditioned on the algorithm's history through iteration $k$, the candidate set $\hat{\mathcal{I}}(\theta_k)$ is fixed, and the samples drawn at iteration $k+1$ are independent of the past. 
Thus, we establish almost-sure termination by showing that, at every iteration, the conditional probability of termination at iteration $k+1$ is uniformly bounded below by a positive constant.
\begin{lemma}[Uniform lower bound on the probability of termination at iteration $k+1$] \label{lemma:stepk1}
Given Assumption \ref{assump:initset} and $r_{\max}<1$, there exists $\delta>0$ independent of $k$ such that,
        \begin{equation}
            \mathbf{P}(T \leq k+1 | T> k) \geq \delta.
        \end{equation}
\end{lemma}
Lemma \ref{lemma:stepk1} establishes that Alg.~\ref{alg:main} has a nonzero probability of terminating at each iteration. We use this result to prove that Alg.~\ref{alg:main} terminates in finite iterations with probability 1.
\begin{theorem}\label{thm:convergence}
    Given Assumption \ref{assump:initset}, Alg.~\ref{alg:main} terminates in a finite number of iterations, i.e., $T < \infty$, with probability 1.
\end{theorem}
The proof of Theorem \ref{thm:convergence} follows by showing that the probability of non-termination decays with the number of iterations, and thus converges to zero as $k\rightarrow \infty$.
Proofs of Lemma \ref{lemma:stepk1} and Theorem \ref{thm:convergence} are provided in Appendix~\ref{app:proofs}, along with practical insights into the convergence behavior of different set representations. Appendix \ref{appendix:discussion} further discusses guidance on selecting $\epsilon_{thresh}$. 



\section{Conclusion}
We present an algorithm that leverages sampling-based optimization to compute discrete-time invariant sets of hybrid systems via their discrete Poincar\'e dynamics.
Our approach requires minimal structural assumptions on the dynamical system and provides probabilistic guarantees on violation of finite-step set invariance.
Additionally, under suitable conditions, finite-time termination is guaranteed with probability $1$.
We demonstrate the algorithm on three dynamical systems, including one where the true invariant set is not explicitly known.
When compared to a baseline method, our algorithm produces less conservative estimates and provides PAC bounds derived from the holdout method.
Future work should explore alternative probabilistic tools for establishing finite-sample guarantees over the full algorithm, including union-bound-based approaches, as well as optimization formulations that account for both invariant and non-invariant samples during fitting.

\section{Acknowledgments}
This work is supported by the National Science Foundation (NSF) CPS Award \#2440387. Elizabeth Dietrich was also supported by an NSF Graduate Research Fellowship.

\bibliography{ref} 

\appendix
\section*{Appendix Overview}
This appendix provides additional information and extended results. Appendix \ref{app:proofs} presents detailed proofs of Lemma~\ref{lemma:stepk1} and Theorem~\ref{thm:convergence}. Appendix \ref{app:results} provides additional experimental results and implementation details, including the computed invariant sets for the Convex Expander-Contractor (CEC), further details on the Compass-Gait Walker, and analysis of multi-step verification and convergence. Finally, Appendix \ref{appendix:discussion} discusses limitations of the proposed approach and provides practical guidance for its application, including recommendations for hyperparameter tuning.
\setcounter{lemma}{0}
\setcounter{theorem}{0}
\section{Proofs}
\label{app:proofs}
\begin{lemma}[Uniform lower bound on the probability of termination at iteration $k+1$] \label{lemma:stepk1}
Given Assumption \ref{assump:initset} and $r_{\max}<1$, there exists $\delta>0$ independent of $k$ such that, 
        \begin{equation}
            \mathbf{P}(T \leq k+1 | T> k) \geq \delta.
        \end{equation}
\begin{proof}
Start with an initial set $\hat{\mathcal{I}}(\theta_0)$ satisfying Assumption~\ref{assump:initset} and apply Alg.~\ref{alg:main}.
Fix an iteration $k$ and let $\mathcal{F}_k$ be Alg.~\ref{alg:main}'s history through $k$. Condition on the event that the algorithm has not yet terminated $\{ T > k\} \in \mathcal{F}_k$.
Then the current candidate set  $\hat{\mathcal{I}}(\theta_k)$ is fixed, and its true escape rate is defined by \eqref{eq:trueescaperate}.
At iteration $k+1$, Alg.~\ref{alg:main} takes the candidate set from the previous iteration, $\hat{\mathcal{I}}(\theta_{k})$, and draws $N$ i.i.d. samples $\{x_i\}^N_{i=1}$ uniformly from $\hat{\mathcal{I}}(\theta_{k})$.
By definition of the true escape rate, each sample violates set containment with probability $r(\hat{\mathcal{I}}(\theta_{k}))$. 
Therefore, given $\{T > k\}$, the probability of observing zero violations is 
\begin{equation}
    \mathbf{P}(| \mathcal{V}_{k}|=0 \mid \mathcal{F}_k ) = (1-r(\hat{\mathcal{I}}(\theta_{k})))^N \geq (1-r_{\max})^N.
\end{equation}
If $| \mathcal{V}_{k}|=0$, then $\epsilon_{k}= 1- \beta^{1/N} \leq \epsilon_{thresh}$
(Remark~\ref{corr:smallestepsilon}), and Alg.~\ref{alg:main} terminates, i.e., $k$ is incremented and $T=k+1$.
Since $\{ T > k\} \in \mathcal{F}_k$, the tower property gives
\begin{align}
    \mathbf{P}(T \leq k+1 \mid T > k ) &\geq \mathbf{P}(| \mathcal{V}_{k}|=0 \mid T > k ) \\
    &= \mathbb{E}[\mathbf{P}(| \mathcal{V}_{k}|=0 \mid \mathcal{F}_k) | T>k]
    \geq  (1-r_{\max})^N =: \delta > 0. \notag
        \end{align}
The second inequality follows from the uniform bound $r_{\max}$ on the true escape rate. Since $\delta$ is independent of $k$, the probability of termination at iteration $k+1$ is uniformly bounded below. 
\end{proof}
\end{lemma}
Lemma \ref{lemma:stepk1} establishes a uniform positive lower bound, which provides a sufficient condition for almost-sure termination. Although an iteration-dependent bound is also possible, it would require additional assumptions on the evolution of $\hat{\mathcal{I}}(\theta_{k})$. The uniform bound, in contrast, relies only on the worst-case escape rate $r_{\max}$. 
In practice, $(1-r_{\max})^N$ is impractically small, unless $r_{\max}$ is small relative to $\frac{1}{N}$. Therefore, if the parameterization $\theta$ cannot represent the true invariant set with an escape rate below $\epsilon_{thresh}$, termination may take a prohibitively long time.
To this end, Theorem \ref{thm:convergence} establishes termination, but does not provide a convergence-rate guarantee. Practical guidance for selecting hyperparameters, including $\epsilon_{thresh}$ and $\theta$, is provided in Appendix \ref{appendix:discussion}.

\begin{theorem}
Given Assumption \ref{assump:initset}, Alg.~\ref{alg:main} terminates in a finite number of iterations, i.e., $T < \infty$, with probability 1.
\begin{proof} 
    The event that Alg.~\ref{alg:main} does not terminate in $k$ iterations is denoted by $\{T>k\}$. 
    Since $\{T>k-1\} \supseteq \{T>k\} $ and $\mathbf{P}(T>0) = 1$, the chain rule gives
        \begin{align}
            \mathbf{P}(T>k) &= \prod^k_{j=1}\mathbf{P}(T > j | T > j-1)\\
            &= \prod^k_{j=1}1 - \mathbf{P}(T \leq j | T > j-1)\\
            &= \prod^{k-1}_{j=0}1 - \mathbf{P}(T \leq j + 1 | T > j) \leq \prod^{k-1}_{j=0}1 - \delta = (1 - \delta)^k,
        \end{align}
    where the last step uses the uniform positive lower bound $\delta$ from Lemma~\ref{lemma:stepk1}.
    
    The event that Alg.~\ref{alg:main} never terminates is $\{ T=\infty\} = \bigcap^\infty_{k=1}\{ T > k\}$.
        Since the events $\{T>k\}$ form a decreasing sequence (i.e., $\{ T > 1\} \supseteq \{ T > 2\} \supseteq  \dots \supseteq \{ T > k\}$) and $\delta>0$, then by \textit{continuity from above}
        \begin{align}
            \mathbf{P}(\{ T=\infty\}) 
            &= \mathbf{P}\left(\bigcap^\infty_{k=1} \{ T>k\}\right)\\
            &= \lim_{k\to \infty} \mathbf{P}(T>k) \\
            &\leq \lim_{k\to \infty} (1-\delta)^k = 0.
        \end{align}
    Since $\mathbf{P}(\{ T=\infty\})\leq0$, Alg.~\ref{alg:main} terminates in finite steps with probability 1, i.e., $\mathbf{P}( T<\infty)=1$.
\end{proof}
\end{theorem}

\section{Additional Results and Implementation Details}\label{app:results}

\subsection{Convex Expander-Contractor (CEC)}
\begin{figure}[H]
    \centering
    \resizebox{\linewidth}{!}{
    \pgfplotsset{compat=1.18}

\newcommand{\figdir}{l4dc_figures/pdfs/}
\newcommand{\pw}{5cm}   
\newcommand{\ph}{5cm}   
\newcommand{\pitch}{9.1cm}

\definecolor{cU}{rgb}{0.692840,0.165141,0.564522}   
\definecolor{cZ}{rgb}{0.915330,0.455146,0.339446}   
\definecolor{cW}{rgb}{0.254627,0.013882,0.615419}   
\definecolor{cV}{rgb}{0.990031,0.730802,0.178361}   

\newcommand{\legline}[1]{\tikz[baseline=-0.5ex]{\draw[#1] (0,0)--(0.7,0);}}
\newcommand{\legdot}[1]{\tikz[baseline=-0.5ex]{\fill[#1] (0,0) circle (2.6pt);}}

\pgfplotsset{
  colormap={plasma}{
    rgb=(0.050383,0.029803,0.527975)
    rgb=(0.254627,0.013882,0.615419)
    rgb=(0.417642,0.000564,0.658390)
    rgb=(0.562738,0.051545,0.641509)
    rgb=(0.692840,0.165141,0.564522)
    rgb=(0.798216,0.280197,0.469538)
    rgb=(0.881443,0.392529,0.383229)
    rgb=(0.949217,0.517763,0.295662)
    rgb=(0.987621,0.652025,0.211364)
    rgb=(0.992440,0.809579,0.145357)
    rgb=(0.940015,0.975158,0.131326)
  },
}

\pgfplotsset{
  panel/.style={
    width=\pw, height=\ph,
    axis on top,
    scale only axis,
    tick align=outside,
    tick style={black},
    every tick label/.append style={font=\normalsize},
    label style={font=\Large},
    xlabel style={yshift=2pt},
    legend style={font=\small, fill=white, fill opacity=0.9, text opacity=1,
                  draw=black!30, cells={anchor=west}},
    legend pos=north east,
    colormap name=plasma,
  },
  zf3/.style={/pgf/number format/.cd, fixed, fixed zerofill, precision=3},
  zf0/.style={/pgf/number format/.cd, fixed, precision=0},
}

\begin{tikzpicture}
\node[anchor=south, draw=black!30, fill=white, inner sep=4pt, font=\Large]
  at (11.6cm, 5.8cm) {%
  \legline{black, dashed, thick}\;$\mathcal{E}_0$\quad
  \legline{black, thick}\;$\mathcal{E}^*$\quad
  \legdot{cU}\;$\mathcal{U}$\quad
  \legdot{cZ}\;$\mathcal{Z}$\quad
  \legdot{cW}\;$\mathcal{W}$\quad
  \legdot{cV}\;$\mathcal{V}$%
};


\begin{axis}[
  panel, at={(0,0)}, anchor=south west, name=p1,
  xmin=-4.1, xmax=4.1, ymin=-3.85, ymax=3.85,
  xtick={-2,0,2}, ytick={-2,0,2},
  xtick pos=left,
  ytick pos=left,
  ylabel style={yshift=-12pt},
  xlabel={$x_1$}, ylabel={$x_2$},
  xticklabel style={/pgf/number format/.cd, fixed, fixed zerofill, precision=1},
  yticklabel style={/pgf/number format/.cd, fixed, fixed zerofill, precision=1},
  colorbar,
  point meta min=1, point meta max=5,
  colorbar style={ytick={1,2,3,4,5}, ylabel={Iteration}, width=0.25cm,
                  yticklabel style={font=\normalsize}, ylabel style={font=\normalsize}},
]
  \addplot graphics[xmin=-4.1,xmax=4.1,ymin=-3.85,ymax=3.85]
    {\figdir Systems.ExpanderContractor_ellipsoid_log_20261110_153217_sets_bare.pdf};
\end{axis}
\node[anchor=north, font=\Large] at ([yshift=-1.3cm]p1.south) {(a)};

\begin{axis}[
  panel, at={(\pitch+.2cm,0)}, anchor=south west, name=p2,
  xmin=-0.6035, xmax=2.602, ymin=-1.267, ymax=3.265,
  xtick pos=left,
  ytick pos=left,
  xtick={0,1,2}, ytick={0,1.5,3.0},
  scaled x ticks=false,
  scaled y ticks=false,
  ylabel style={yshift=-7pt},
  xticklabel style={/pgf/number format/.cd, fixed, fixed zerofill, precision=1},
  yticklabel style={/pgf/number format/.cd, fixed, fixed zerofill, precision=1},
  xlabel={$x_1$}, ylabel={$x_2$},
]
  \addplot graphics[xmin=-0.6035,xmax=2.602,ymin=-1.267,ymax=3.265]
    {\figdir Systems.ExpanderContractor_ellipsoid_log_20261110_153217_invariant_bare.pdf};
\end{axis}
\node[anchor=north, font=\Large] at ([yshift=-1.3cm]p2.south) {(b)};

\begin{axis}[
  panel, at={(2*\pitch,0)}, anchor=south west, name=p3,
  xmin=-2.9655, xmax=3.3858, ymin=-2.8111, ymax=3.5403,
  xtick pos=left,
  ytick pos=left,
  xtick={-2,0, 2}, ytick={-2,0,2},
  scaled x ticks=false,
  scaled y ticks=false,
  ylabel style={yshift=-7pt},
  xticklabel style={/pgf/number format/.cd, fixed, fixed zerofill, precision=1},
  yticklabel style={/pgf/number format/.cd, fixed, fixed zerofill, precision=1},
  xlabel={$x_1$}, ylabel={$x_2$},
  colorbar,
  point meta min=1, point meta max=9,
  colorbar style={ytick={2, 4, 6, 8, 10}, ylabel={Iteration}, width=0.25cm,
                  yticklabel style={font=\normalsize}, ylabel style={font=\normalsize}},
]
  \addplot graphics[xmin=-2.9655, xmax=3.3858, ymin=-2.8111, ymax=3.5403]
    {\figdir ExpanderContractor_rbf_progression_bare.pdf};
\end{axis}
\node[anchor=north, font=\Large] at ([yshift=-1.3cm]p3.south) {(c)};

\end{tikzpicture}%
    }
    \caption{(a) Sequence of ellipsoid iterates $\hat{\mathcal{E}}(\theta_0), \hat{\mathcal{E}}(\theta_1), \ldots, \hat{\mathcal{E}}(\theta^*)$  produced by the algorithm for the Convex Expander-Contractor, with accuracy threshold $\epsilon_{thresh}=1.031\times10^{-2}$. (b) Verification of the final set $\hat{\mathcal{E}}(\theta^*)$, showing the sampled preimages $\mathcal{U}$ that return to $\hat{\mathcal{E}}(\theta^*)$ and the escaping preimages $\mathcal{Z}$, together with their respective Poincaré return images $\mathcal{W}$ and $\mathcal{V}$. (c) Sequence of RBF iterates $\hat{\mathcal{R}}(\theta_0), \hat{\mathcal{R}}(\theta_1), \ldots, \hat{\mathcal{R}}(\theta^*)$, with accuracy threshold $\epsilon_{thresh}=0.06$. }
    \label{fig:cecsolution}
\end{figure}

We evaluate Alg.~\ref{alg:main} over 10 trials using different random seeds with ellipsoids. 
On average across 10 trials with 10 random seeds, the algorithm recovers an invariant ellipsoid with $\epsilon^*=0.0111$, in 7.5 iterations.
Fig. \ref{fig:cecsolution} illustrates the progression of the ellipsoid iterates in one representative trial and the resulting verification of the final set $\hat{\mathcal{E}}(\theta^*)$. The volume of the recovered invariant ellipsoid is only 1.1\% smaller than that of the true invariant set. When Alg.~\ref{alg:main} is instead run with 3 RBFs and $\epsilon_{thresh}=0.06$ across 10 trials, it achieves an average $\epsilon^*=0.054$ after 32.2 iterations. The recovered invariant set closely resembles an ellipsoid but requires substantially more iterations to converge, reflecting the poor fit of a nonconvex representation to an inherently ellipsoidal set. One representative run is shown in Fig. \ref{fig:cecsolution}. 

\subsection{Compass-Gait Walker}
\subsubsection{Dynamics and Poincar\'e Map}
We provide additional details on the Compass-Gait Walker, including its hybrid dynamics and the numerical construction of the associated Poincar\'e return map.  

We define the state vector for the continuous dynamics as $x_c=[q^T, \dot{q}^T]^T$, where $q=[\theta_{sw}, \theta_{st}]^T$ denotes the swing and stance leg angles, measured relative to vertical.
The continuous dynamics are given by the manipulator equation $H(q)\ddot{q} + C(q, \dot{q})\dot{q} + G(q)=0$, where $H$, $C$, and $G$ denote the inertia, Coriolis, and gravitational terms, respectively.
When walking down a slope with angle $\gamma$, the foot hits the ground and a reset is triggered when $\theta_{sw}+\theta_{st}-2\gamma = 0$.
Thus, the guard surface is defined as $\mathcal{S}=\left\{x:h(x)=\theta_{sw}+\theta_{st}-2\gamma = 0 \right\}$.
The reset map dynamics follow from the inelastic collision between the swing leg and ground, $Q^+(q^+)\dot{q}^+ = Q^-(q^-)\dot{q}^-$.
The matrices $H, C, G, Q^+, Q^-$ defining these dynamics are provided below \citep{manchester2010regionsattractionhybridlimit}:
\begin{align*}
H &= \begin{bmatrix}
mb^2 & -mlb\cos(\theta_{st}-\theta_{sw}) \\
-mlb\cos(\theta_{st}-\theta_{sw}) & (m_h+m)l^2 + ma^2
\end{bmatrix}, \\[1ex]
C &= \begin{bmatrix}
0 & mlb\sin(\theta_{st}-\theta_{sw})\dot{\theta}_{st} \\
-mlb\sin(\theta_{st}-\theta_{sw})\dot{\theta}_{sw} & 0
\end{bmatrix}, \quad
G = \begin{bmatrix}
mbg\sin(\theta_{sw}) \\
-(m_h l + ma + ml)g\sin(\theta_{st})
\end{bmatrix}, \\[1ex]
Q^+ &= \begin{bmatrix}
mb\big(b - l\cos(2\alpha)\big) & ml\big(l - b\cos(2\alpha)\big) + ma^2 + m_h l^2 \\
mb^2 & -mbl\cos(2\alpha)
\end{bmatrix}, \\[1ex]
Q^- &= \begin{bmatrix}
-mab & -mab + (m_h l^2 + 2mal)\cos(2\alpha) \\
0 & -mab
\end{bmatrix}.
\end{align*}
In our experiments, we set the leg mass to $m$=5, hip mass to $m_h$=10, the leg center-of-mass distances to $a=b=0.5$, and the slope to $\gamma=0.0525$.

We apply the approach from Sec. \ref{sec:HybridSystems} to derive a Poincar\'e map 
$\mathcal{P}:\mathbb{R}^3\rightarrow \mathbb{R}^3$,
which must be computed numerically using \texttt{python}.
To represent the Poincar\'e map, we first define coordinate transformations from the continuous coordinates $x_c \in \mathbb{R}^4$ to a local set of coordinates on the guard surface $x\in S$.
Let $x=B(x_c-x_c^*)$ and $x_c=B^{\dagger}x + x_c^*$, where $x_c^*$ is the fixed point of the hybrid system represented in the continuous dynamics coordinates, and the matrices  $B, B^{\dagger}$ are defined as:
\begin{align*}
B=
\begin{bmatrix}
0.5 & -0.5 & 0 & 0 \\
0 & 0 & 1 & 0 \\
0 & 0 & 0 & 1 \\
\end{bmatrix}, \quad 
B^{\dagger} = 
\begin{bmatrix}
1 & 0 & 0 \\
-1 & 0 & 0 \\
0 & 1 & 0 \\
0 & 0 & 1 \\
\end{bmatrix}.
\end{align*}

The Poincar\'e map is a function that takes a point $x \in S$, converts it to the continuous coordinates $x_c$, applies the reset map, numerically integrates the continuous dynamics until the trajectory intersects the guard, and maps this intersection back to the lower-dimensional coordinates.
We implement this procedure using the event-based integration capabilities of the differential equation solver \texttt{diffrax} \citep{kidger2021on}, built on the accelerator library \texttt{jax} \citep{jax2018github}, enabling parallel evaluation of $\mathcal{P}$. Further implementation details are available in our accompanying GitHub repository.

\subsubsection{Baseline Implementations}



\citep{madabushi2026} proposed an approach for computing forward-invariant sets of hybrid systems by constructing a differential inclusion that overapproximates the error dynamics of the system around a nominal periodic orbit.
This method constructs an overapproximating tube around the continuous phase of the periodic orbit and computes its intersections with the guard surface.
Forward invariance is established by verifying that all intersections remain within the tube after application of the reset map.
While this approach provides a formal certificate of invariance, it can be highly conservative.
In particular, the use of differential inclusions to overapproximate both the continuous dynamics and reset map can introduce significant conservatism.
Consequently, the resulting forward-invariant set is substantially smaller than that recovered by Alg. \ref{alg:main}, which operates directly on samples from high-fidelity dynamics rollouts.
We did not compare to the sphere-based approach of \citep{tucker2024synthesizing}, because in general, a sphere is not guaranteed to be forward-invariant, even for a system with an asymptotically stable linearization.
This is the case for the Compass-Gait Walker, where the sphere-based approach failed to find an invariant set.


\subsection{Multi-Step Verification }
\label{app:verification}
\begin{figure}[H]
    \centering
    \resizebox{.5\linewidth}{!}{
    \definecolor{CaliforniaGold}{rgb}{0.992, 0.71, 0.082}
\definecolor{darkblue}{rgb}{0.0, 0.13, 0.28}
\definecolor{darkred}{rgb}{0.55, 0.0, 0.0}
\definecolor{darkcyan}{rgb}{0.0, 0.55, 0.55}
\definecolor{deepcarrotorange}{rgb}{0.54, 0.2, 0.14}
\definecolor{navyblue}{rgb}{0.0, 0.0, 0.5}
\definecolor{rose}{rgb}{1.0, 0.0, 0.5}
\definecolor{amber}{rgb}{1.0, 0.49, 0.0}

\usepgfplotslibrary{fillbetween}
\usetikzlibrary{patterns}

\pgfplotsset{compat=1.16}
\hspace{-2cm}
\begin{tikzpicture}
        \begin{axis}[
        width=12cm,
        height=6cm,
        xlabel={$n$ steps},
        ylabel={$1-\epsilon$},
        ymin=0.85, ymax=1.0,
        scaled y ticks=false,
        yticklabel={\pgfmathprintnumber[fixed, precision=2]{\tick}},
        grid=both,
        grid style={line width=0.2pt, draw=gray!30},
        major grid style={line width=0.4pt, draw=gray!60},
        tick align=outside,
        axis lines=left,
        axis line style={-},
        mark size=2pt,
        thick,
        legend style={
          at={(0.725, 0.225)},
          anchor=south west,
          draw=gray,
          rounded corners=3pt,
          font=\small,
        },
        legend entries={Compass-Gait, CEC, NEC},
      ]
 
      \addplot[
      color=rose!70,
      smooth,
      line width=2pt,
    ] table [
      col sep=comma,
      x=n_steps,
      y expr={1 - \thisrow{epsilon}},
    ] {fig/data/epsilon_ksteps.csv};

    \addplot[
      color=amber!70,
      smooth,
      line width=2pt,
    ] table [
      col sep=comma,
      x=n_steps,
      y expr={1 - \thisrow{epsilon}},
    ] {fig/data/expander_epsilon_ksteps.csv};

    \addplot[
      color=violet!70,
      smooth,
      line width=2pt,
    ] table [
      col sep=comma,
      x=n_steps,
      y expr={1 - \thisrow{epsilon}},
    ] {fig/data/dogbone_epsilon_ksteps.csv};
    
    \end{axis}
    \end{tikzpicture}
    }
    \caption{Evaluation of extension of one-step probabilistic guarantees to $n$-step invariance. $\epsilon$ is evaluated over trajectories propagated for $n$ steps: stable probabilities suggest the one-step guarantee of Alg. \ref{alg:main} is robust.
    }    
    \label{fig:bounds}
\end{figure}
We evaluated the accuracy of the recovered ellipsoidal invariant sets over $n=1,\dots,20$ successive applications of the Poincar\'e map for all three examples: CEC, NEC, and the Compass-Gait Walker. For each value of $n$, we construct a test set by sampling $N$ points from $\hat{\mathcal{E}}(\theta^*)$ and evaluating whether their images under $n$ successive applications of $\mathcal{P}$ remain within $\hat{\mathcal{E}}(\theta^*)$. This does not require the point to remain inside $\mathcal{E}(\theta^*)$ at intermediate steps.
As shown in Fig. \ref{fig:bounds}, the observed violation probability remains relatively stable as $n$ increases, despite the probabilistic guarantees being derived specifically for one-step invariance.
This suggests that, in practice, the one-step probabilistic guarantees can extend to longer horizons without substantial degradation. However, these results do not investigate trajectory-wise invariance over the entire horizon, which would be an interesting direction for future work.

For the NEC example, however, Fig. \ref{fig:bounds} shows a decrease in accuracy beyond $n=2$ steps. This highlights the limitations of the ellipsoidal representation for this system, suggesting that some points contained in the estimated one-step ellipsoidal invariant set do not remain invariant under repeated applications of $\mathcal{P}$.
In contrast, for the Compass-Gait Walker, the observed violation probability decreases with increasing $n$ and remains low across all tested horizons. This indicates that the estimated ellipsoid provides a conservative approximation of the invariant set, despite the true invariant set being unknown. 

\subsection{Evaluation of Convergence}
\label{app:convergence}
Fig.~\ref{fig:convergence} illustrates the convergence behavior of Alg.~\ref{alg:main} for ellipsoidal invariant sets across the three examples, showing the evolution of the accuracy measure $1-\epsilon$ and the volume of the set over successive iterations. We report the mean and standard deviation over $10$ independent runs. In all three examples, the accuracy increases towards the user-defined threshold, while the volume of the estimated invariant set decreases. These results demonstrate that the proposed algorithm achieves the desired probabilistic accuracy across systems with varying dynamics and invariant set representations, although the number of iterations required varies considerably.
\begin{figure}[H]
    \centering
    \resizebox{\linewidth}{!}{

\definecolor{deepcarrotorange}{rgb}{0.54, 0.2, 0.14}
\definecolor{rose}{rgb}{0.65, 0.04, 0.37}
\definecolor{amber}{rgb}{1.0, 0.49, 0.0}
\definecolor{cerise}{rgb}{1.0, 0.11, 0.81}
\definecolor{lavender}{rgb}{0.67, 0.38, 0.8}
\definecolor{lightpurple}{rgb}{0.4, 0.01, 0.24}
\definecolor{sunglow}{rgb}{0.66, 0.66, 0.66}

\pgfplotsset{
  compat=1.16,
  every axis/.append style={
    scale only axis,
    xlabel style={font=\footnotesize},
    ylabel style={font=\footnotesize},
    title style={font=\footnotesize, yshift=-1mm},
    xticklabel style={font=\footnotesize},
    yticklabel style={font=\footnotesize},
    xtick align=outside,
    ytick align=outside,
    axis line style={-},
  }
}

\newcommand{\plotwidth}{3.2cm}
\newcommand{\plotheight}{2.8cm}
\newcommand{\panelstep}{5.0cm} 

\begin{tikzpicture}

  \begin{axis}[
    name=main1, at={(0,0)}, anchor=south west,
    width=\plotwidth, height=\plotheight,
    title={Convex Expander-Contractor},
    xlabel={$k$ Iterations},
    xlabel style={yshift=-1mm},
    ylabel={$1-\epsilon_k$},
    ylabel style={color=black, yshift=1mm},
    yticklabel style={color=deepcarrotorange},
    ytick style={color=deepcarrotorange},
    xmin=0, xmax=6, xtick={0, 2, 4, 6},
    ymin=0.1, ymax=1
  ]
    \addplot[name path=aeps_upper, deepcarrotorange!30, forget plot]
      table [col sep=comma, x=iter, y=upper] {l4dc_figures/data/cec_epsilon_summary.csv};
    \addplot[name path=aeps_lower, deepcarrotorange!30, forget plot]
      table [col sep=comma, x=iter, y=lower] {l4dc_figures/data/cec_epsilon_summary.csv};
    \addplot[deepcarrotorange!30, forget plot] fill between[of=aeps_upper and aeps_lower];
    \addplot[deepcarrotorange, thick, mark=none]
      table [col sep=comma, x=iter, y=mean] {l4dc_figures/data/cec_epsilon_summary.csv};
    \addplot[sunglow, dashed, ultra thick, domain=0:6, samples=2, forget plot] {1-0.0103};
  \end{axis}
  \begin{axis}[
    at={(main1.south west)}, anchor=south west,
    width=\plotwidth, height=\plotheight,
    xmin=0, xmax=6, xtick={1, 2, 3, 4, 5, 6},
    axis x line=none,
    axis y line*=right,
    yticklabel style={color=amber},
    ytick style={color=amber},
  ]
    \addplot[name path=avol_upper, amber!30, forget plot]
      table [col sep=comma, x=iter, y=upper] {l4dc_figures/data/cec_volume_summary.csv};
    \addplot[name path=avol_lower, amber!30, forget plot]
      table [col sep=comma, x=iter, y=lower] {l4dc_figures/data/cec_volume_summary.csv};
    \addplot[amber!30, forget plot] fill between[of=avol_upper and avol_lower];
    \addplot[amber, thick, mark=none]
      table [col sep=comma, x=iter, y=mean] {l4dc_figures/data/cec_volume_summary.csv};
  \end{axis}

  \begin{axis}[
    name=main2,
    at={(\panelstep,0)}, anchor=south west,
    width=\plotwidth, height=\plotheight,
    title={Nonconvex Expander-Contractor},
    xlabel={$k$ Iterations},
    xlabel style={yshift=-1mm},
    yticklabel style={color=lightpurple},
    ytick style={color=lightpurple},
    xmin=0, xmax=118,
    ymin=0.7, ymax=1.0,
    ytick={0.7, 0.8, 0.9, 1.0},
  ]
    \addplot[name path=beps_upper, lightpurple!30, forget plot]
      table [col sep=comma, x=iter, y=upper] {l4dc_figures/data/nec_epsilon_summary.csv};
    \addplot[name path=beps_lower, lightpurple!30, forget plot]
      table [col sep=comma, x=iter, y=lower] {l4dc_figures/data/nec_epsilon_summary.csv};
    \addplot[lightpurple!30, forget plot] fill between[of=beps_upper and beps_lower];
    \addplot[lightpurple, thick, mark=none]
      table [col sep=comma, x=iter, y=mean] {l4dc_figures/data/nec_epsilon_summary.csv};
    \addplot[sunglow, dashed, ultra thick, domain=0:118, samples=2, forget plot] {1-0.07};
  \end{axis}
  \begin{axis}[
    at={(main2.south west)}, anchor=south west,
    width=\plotwidth, height=\plotheight,
    xmin=0, xmax=118,
    ymin=1.5, ymax=3.5,
    axis x line=none,
    axis y line*=right,
    yticklabel style={color=lavender},
    ytick style={color=lavender},
  ]
    \addplot[name path=bvol_upper, lavender!30, forget plot]
      table [col sep=comma, x=iter, y=upper] {l4dc_figures/data/nec_volume_summary.csv};
    \addplot[name path=bvol_lower, lavender!30, forget plot]
      table [col sep=comma, x=iter, y=lower] {l4dc_figures/data/nec_volume_summary.csv};
    \addplot[lavender!30, forget plot] fill between[of=bvol_upper and bvol_lower];
    \addplot[lavender, thick, mark=none]
      table [col sep=comma, x=iter, y=mean] {l4dc_figures/data/nec_volume_summary.csv};
  \end{axis}

  \begin{axis}[
    name=main3,
    at={(2*\panelstep,0)}, anchor=south west,
    width=\plotwidth, height=\plotheight,
    title={Compass-Gait Walker},
    xlabel={$k$ Iterations},
    xlabel style={yshift=-1mm},
    yticklabel style={color=rose},
    ytick style={color=rose},
    ytick={0.6, 0.8, 1.0},
    xtick={0, 10, 20, 30, 40},
    xmin=0, xmax=49,
    ymin=0.75, ymax=1,
    ytick={0.8, 0.9, 1.0},
  ]
    \addplot[name path=ceps_upper, rose!30, forget plot]
      table [col sep=comma, x=iter, y=upper] {l4dc_figures/data/cgw_epsilon_summary.csv};
    \addplot[name path=ceps_lower, rose!30, forget plot]
      table [col sep=comma, x=iter, y=lower] {l4dc_figures/data/cgw_epsilon_summary.csv};
    \addplot[rose!30, forget plot] fill between[of=ceps_upper and ceps_lower];
    \addplot[rose, thick, mark=none]
      table [col sep=comma, x=iter, y=mean] {l4dc_figures/data/cgw_epsilon_summary.csv};
    \addplot[sunglow, dashed, ultra thick, domain=1:74, samples=2, forget plot] {1-0.04};
  \end{axis}
  \begin{axis}[
    at={(main3.south west)}, anchor=south west,
    width=\plotwidth, height=\plotheight,
    xmin=0, xmax=49,
    axis x line=none,
    axis y line*=right,
    scaled y ticks=false,
    ymin=0.001, ymax=0.0035,
    ytick={0.001, 0.002, 0.003, 0.004},
    yticklabel={\pgfmathprintnumber[fixed, precision=4]{\tick}},
    ylabel={$\mathrm{Vol}(\hat{\mathcal{E}}(\theta))$},
    ylabel style={color=black, yshift=-1mm},
    yticklabel style={color=cerise},
    ytick style={color=cerise},
  ]
    \addplot[name path=cvol_upper, cerise!30, forget plot]
      table [col sep=comma, x=iter, y=upper] {l4dc_figures/data/cgw_volume_summary.csv};
    \addplot[name path=cvol_lower, cerise!30, forget plot]
      table [col sep=comma, x=iter, y=lower] {l4dc_figures/data/cgw_volume_summary.csv};
    \addplot[cerise!30, forget plot] fill between[of=cvol_upper and cvol_lower];
    \addplot[cerise, thick, mark=none]
      table [col sep=comma, x=iter, y=mean] {l4dc_figures/data/cgw_volume_summary.csv};
  \end{axis}

\end{tikzpicture}
    }
    \caption{Mean and standard deviation of accuracy $1-\epsilon$ and volume $\text{Vol}(\hat{\mathcal{E}}(\theta))$ over 10 algorithm runs using ellipsoids. The x-axis is truncated for the shortest run.
    The dashed gray line indicates the user-defined target accuracy for each system.
    (Left): The Convex Expander-Contractor system (accuracy: dark orange, volume: light orange) with desired accuracy $98.97\%$, converges within 7.5 iterations on average.
    (Middle): The Nonconvex Expander-Contractor system (accuracy: dark purple, volume: light purple) with desired accuracy $93\%$, converges within 138.6 iterations on average.
    (Right): The Compass-Gait Walker system (accuracy: dark pink, volume: light pink) with desired accuracy $96\%$, converges within 63.8 iterations on average.
    }
    \label{fig:convergence}
\end{figure}


\section{Discussion}\label{appendix:discussion}
The invariant set identification problem lies at the intersection of formal guarantees and practical applicability. We discuss three aspects of this intersection: the conditions under which finite-step invariant sets exist, the limitations of sampling-based approaches, and practical considerations for tuning hyperparameters of the algorithm.

\subsection{When Does a Finite-Step Invariant Set Exist?}
The proposed algorithm presupposes the existence of a finite-step invariant set; therefore, it is worth examining when this assumption is well-founded.
In general, the existence of such a set is not guaranteed; a system may not admit a compact invariant set in the step-to-step dynamics.
A verifiable sufficient condition is the existence of a stable fixed point on the Poincar\'e surface. 
By discrete-Lyapunov theory, a stable fixed point implies the existence of a corresponding invariant set in the step-to-step dynamics \citep{bof2018lyapunovtheorydiscretetime}, 
which can then be characterized in shape and volume using Alg.~\ref{alg:main}.
This condition is broadly applicable to hybrid systems admitting stable periodic orbits, including bipedal locomotion controllers synthesized to track a specific gait reference \citep{tucker2024synthesizing}.
Beyond analytically verifiable conditions, the sampling-based nature of our approach permits a more pragmatic perspective: convergence of the algorithm serves as empirical evidence for the existence of a finite-step invariant set within the region of interest.
This is consistent with the broader spirit of the framework, where PAC-style guarantees validate the returned set rather than relying on a priori knowledge of the system's structure.

\subsection{Curse of Dimensionality}

Our approach is subject to the curse of dimensionality inherent to sampling-based approaches. Specifically, it requires uniform sampling over the volume of each candidate invariant set and relies on these samples to adequately characterize the system dynamics and capture the true invariant set.
In high-dimensional spaces, more samples are required to fully characterize the volume of this set, posing potential scalability challenges.
This problem can be partially mitigated through large-scale sampling, leveraging parallel processing on Graphics Processing Units (GPUs).
Such approaches have gained traction in the robotics and machine learning communities, with infrastructure such as Brax \citep{brax2021github}, MuJoCo \citep{todorov2012mujoco}, and Isaac Gym \citep{makoviychuk2021isaac}, enabling large-scale physics simulation in contact-rich environments.

\subsection{Tuning Hyperparameters}
To apply Alg. \ref{alg:main}, the user must specify an initial set $\hat{\mathcal{I}}(\theta_0)$ and termination threshold $\epsilon_{thresh}$.
The choice of initial set can be guided by prior knowledge of the system under analysis.
For example, if the system is known to contain a single stable equilibrium point, an ellipsoidal set may be a suitable initial choice, as demonstrated in Sec.~\ref{sec:compass_gait}.
In contrast, systems with multiple stable equilibria may exhibit nonconvex forward-invariant sets, for which a more flexible representation, such as an RBF, may be better suited.
Since the proposed framework makes few assumptions on the system dynamics, selecting an appropriate set representation may require some experimentation. We therefore recommend starting with an ellipsoidal representation, as its fitting procedure in Step 4 of Alg. \ref{alg:main} is computationally the fastest, and exploring alternative representations if the ellipsoidal representation yields unsatisfactory results.

Similarly, the termination threshold $\epsilon_{thresh}$ must be selected for each system and set representation to balance accuracy and convergence time.
Although Alg. \ref{alg:main} is guaranteed to terminate in finite time under sufficient conditions, choosing $\epsilon_{thresh}$ too small can result in prohibitively long convergence times.
In this work, we select $\epsilon_{thresh}$ to be the smallest value that still permits reliable convergence in less than 200 steps.
If fast convergence requires a relatively large $\epsilon_{thresh}$, this may indicate that a different set representation is more appropriate for the system at hand.

\end{document}